\documentclass[aps,english,preprint,nofootinbib]{revtex4}

\usepackage{graphicx}
\usepackage{hyperref}
\usepackage{amsmath, amssymb}
\usepackage{babel}
\usepackage{color}
\usepackage{slashed}
\usepackage{simplewick}

\def\slashchar#1{\setbox0=\hbox{$#1$}           
	\dimen0=\wd0                                    
	\setbox1=\hbox{/} \dimen1=\wd1                  
	\ifdim\dimen0>\dimen1                           
	\rlap{\hbox to \dimen0{\hfil/\hfil}}            
	#1                                             
	\else                                          
	\rlap{\hbox to \dimen1{\hfil$#1$\hfil}}        
	/                                           
	\fi}

\begin{document}

\title{Effective Field Theory in The Study of Long Range Nuclear Parity Violation on Lattice}
 
 
\author{Feng-Kun Guo$^{1,2}$ and Chien-Yeah Seng$^3$}

\affiliation{ $^1$CAS Key Laboratory of Theoretical Physics,
		Institute of Theoretical Physics,\\ Chinese Academy of Sciences,
		Beijing 100190, China\\
$^2$School of Physical Sciences,
University of Chinese Academy of Sciences,
Beijing 100049, China \\
$^3$INPAC, Shanghai Key Laboratory for Particle Physics and Cosmology, \\
MOE Key Laboratory
for Particle Physics, Astrophysics and Cosmology,  \\
School of Physics and Astronomy, Shanghai Jiao-Tong University, Shanghai 200240, China}
  
\date{\today}

 \begin{abstract}
{\color{black}A non-zero signal $A_\gamma^\mathrm{np}=(-3.0\pm1.4\pm0.2)\times 10^{-8}$ of the gamma-ray asymmetry in the neutron-proton capture was recently reported by the NPDGamma Collaboration which provides the first determination of the $\Delta I=1$ parity-odd pion-nucleon coupling constant $h_\pi^1=(2.6\pm 1.2\pm 0.2)\times 10^{-7}$}. The ability to reproduce this value from first principles serves as a direct test of our current understanding of the interplay between the strong and weak interaction at low energy. To motivate new lattice studies of $h_\pi^1$, we review the current status of the theoretical understanding of this coupling, which includes our recent work that relates it to a nucleon mass-splitting by a soft-pion theorem. We further investigate the possibility of calculating the mass-splitting on the lattice by providing effective field theory parameterizations of all the involved quark contraction diagrams. We show that the lattice calculations of the easier connected diagrams will provide information of the chiral logarithms in the much harder quark loop diagrams and thus help in the chiral extrapolation of the latter. 
 \end{abstract}

\maketitle


\section{Introduction}

The study of parity~(P)-violation in nuclear and atomic systems has continued to be a central topic in the low-energy community despite that the P-violation in Standard Model~(SM) electroweak~(EW) sector is well-established and all the EW parameters are already quite precisely measured. 
The reason is that we are really using the hadronic weak interaction~(HWI) as a tool to understand the peculiarities in the strong interaction dynamics. The non-perturbative nature of Quantum Chromodynamics~(QCD) in the confinement region resembles a black box that asserts non-predictable dressings to the confined quarks in a hadron. Therefore, in order to examine its properties, the ``bare" weak interaction which is well-understood serves as a probe inserted into the black box which then returns the HWI that is experimentally measured. The role of P-violation in this procedure is also obvious: as the effective strong interaction coupling is $10^6$ times larger than the weak coupling, one relies entirely on a symmetry-violating signal to disentangle the HWI from the huge strong interaction background. It is therefore not the discovery of a non-zero P-violation signal in HWI, but the precise measurement of its value that will provide us with 
the opportunity of testing the SM with the interplay between weak and strong interactions.

Effects of the hadronic parity violation (HPV) are usually classified according to their isospin, and among all others the $\Delta I=1$ HPV possesses a special role as a unique probe of the hadronic weak neutral current. Moreover, it is the only channel that allows for a single pion-exchange, and hence plays a dominant role in the long-range HPV. Also, the $\Delta I=1$ P-odd pion-nucleon coupling plays a nontrivial role in the $\vec p p$ scattering through two-pion exchange as discussed in Refs.~\cite{deVries:2013fxa,deVries:2014vqa}. The recent observation of a P-violating 2.2~MeV gamma-ray asymmetry {\color{black}$A_\gamma^\mathrm{np}=(-3.0\pm1.4\pm0.2)\times 10^{-8}$} in the polarized neutron capture on hydrogen by the NPDGamma Collaboration~\cite{Blyth:2018aon} provides the first solid experimental confirmation of the isovector HPV, and is promised to create a new stir to the field that has been suffering from a ``slow pace of (experimental) results since 1980"~\cite{Haxton:2013aca}. It is therefore timely to review our current knowledge of HPV and discuss how it could be improved by making the fullest use of the new experimental result.

Early attempts to describe HPV at the phenomenological level are based on isospin symmetry and perturbative expansions of small interaction energies, a strategy that is now inherited by the effective field theory~(EFT) approach. A well-known example of such a kind is the work by Danilov~\cite{Danilov:1965} in the 60s that parameterized the P-odd nucleon-nucleon interaction at very low energy in terms of five $S$-$P$ transition amplitudes with $\Delta I=0,1,2$. The ground-breaking work by Desplanques, Donoghue and Holstein~(DDH)~\cite{Desplanques:1979hn} in the early 80s adopted a very different starting point, namely to describe HPV through single exchange of light mesons $\pi,\rho$ and $\omega$ with seven independent nucleon-meson coupling constants. Despite being a model, its succinctness has attracted much attentions and has become the basis of many experimental analysis. The development of the EFT description of HPV~\cite{Kaplan:1992vj,Kaplan:1998xi,Zhu:2004vw,Girlanda:2008ts,Griesshammer:2010nd,Schindler:2013yua,Viviani:2014zha,deVries:2015gea} signifies a switch to a model-independent framework that features pion-exchanges and contact terms, where a systematic power expansion with respect to a typical small momentum scale $p$ ensures the finiteness of the number of operators needed in any given order. Translation tables, sometimes known informally as the ``Rosetta stone"~\cite{Haxton:2013aca,Gardner:2017xyl}, are available to connect these many different effective descriptions of the same physics~{\color{black}\cite{deVries:2014vqa,Vanasse:2011nd,Schindler:2013yua}} (where the cutoff dependence is also discussed for the translation). Finally, nuclear model calculations have been carried out to connect the HPV coupling strengths to the experimental observables in nuclear or atomic systems; examples in the $\Delta I=1$ channel include Refs.~\cite{Schindler:2013yua,Hyun:2004xp,Liu:2006dm},

It is apparent that none of the frameworks above allows by itself a quantitative connection to the SM EW sector. The latter requires predictions of the theory parameters (such as the Danilov amplitudes, the nucleon-meson couplings in the DDH formalism and the low-energy constants (LECs) in the EFTs) in term of the SM EW parameters, which in turn require a precise control of strong dynamics in the confinement regime. In the original DDH paper, the SU(6) quark model was used to predict a ``reasonable range" $(0-11)\times 10^{-7}$ and the ``best guess" around $4.6\times 10^{-7}$ for the $\Delta I=1$ P-odd pion-nucleon coupling $h_\pi^1$. Subsequent efforts include the use of quark models~\cite{Dubovik:1986pj,Feldman:1991tj,Hyun:2016ddn}, Skyrme models~\cite{Kaiser:1988bt,Kaiser:1989fd,Meissner:1998pu} and QCD sum rules~\cite{Henley:1995ad,Lobov:2002xb}. Their predictions of $h_\pi^1$, together with the NPDGamma outcome, which could be improved over using the analysis of Ref.~\cite{deVries:2015pza}, are summarized in Table~\ref{tab:models}, and one can see that there is in general no agreement between different model predictions.   {\color{black}Recent analyses based on large-$N_c$~\cite{Phillips:2014kna,Schindler:2015nga,Samart:2016ufg,Gardner:2017xyl} suggest a suppression of $h_\pi^1$ from the na\"{\i}ve dimensional analysis result, or in more general terms, a hierarchical structure of the five Danilov amplitudes.} 

\begin{table}[tb]
\caption{\label{tab:models}Existing model calculations of $h_{\pi}^{1}$ in comparison to the {\color{black}implied value from the NPDGamma experiment quoted in Ref. \cite{Blyth:2018aon}}.}

	\begin{centering}
		\begin{tabular}{|l|c|}
			\hline 
			Models & $h_{\pi}^{1}$\tabularnewline
			\hline 
			\hline 
			DDH range \cite{Desplanques:1979hn} & $(0-1)\times10^{-6}$\tabularnewline
			\hline 
			Quark model \cite{Dubovik:1986pj} & $1.3\times10^{-7}$\tabularnewline
			\hline 
			Quark model \cite{Feldman:1991tj} & $2.7\times10^{-7}$\tabularnewline
			\hline 
			Quark model \cite{Hyun:2016ddn}& $8.7\times10^{-8}$\tabularnewline
			\hline 
			SU(2) Skyrme \cite{Kaiser:1988bt} & $1.8\times10^{-8}$\tabularnewline
			\hline 
			SU(2) Skyrme \cite{Kaiser:1989fd}& $2\times10^{-8}$\tabularnewline
			\hline 
			SU(3) Skyrme \cite{Meissner:1998pu}& $(0.8-1.3)\times10^{-7}$\tabularnewline
			\hline 
			QCD sum rule \cite{Henley:1995ad} & $3\times10^{-7}$\tabularnewline
			\hline 
			QCD sum rule \cite{Lobov:2002xb}& $3.4\times10^{-7}$\tabularnewline
			\hline
			{\color{black}NPDGamma} \cite{Blyth:2018aon}& $(2.6\pm 1.2\pm 0.2)\times10^{-7}$\tabularnewline
			\hline 
		\end{tabular}
		\par\end{centering}
\end{table}

Lattice QCD is currently the only available approach to compute low-energy hadronic observables from the first principle with a controlled error. Unfortunately, in contrast to the steady progress made in the lattice calculation of $\Delta I=2$ P-odd amplitudes~\cite{Kurth:2015cvl,WalkerLoud}, there is so-far only one very preliminary study of $h_\pi^1$ by Wasem in Ref.~\cite{Wasem:2011zz} with no follow-ups. In that work, a three-point correlation function is computed to obtain the matrix element $\left\langle n\pi^+\right|\mathcal{O}_{\mathrm{PV}}^{\Delta I=1}\left|p\right\rangle$ with $L=2.5$~fm, $a=0.123$~fm and $m_\pi=389$~MeV, and the reported result is $h_\pi^1=\left(1.099\pm0.505^{+0.058}_{-0.064}\right)\times 10^{-7}$. Despite being consistent with the NPDGamma result, this number should not be taken seriously due to the existence of several unquantified assumptions as pointed out in Ref.~\cite{WalkerLoud}: (1) the three-quark representation of the $N\pi$ interpolator; (2) the negligence of the so-called ``quark loop diagrams"; (3) the calculation was done with only a single choice of volume, lattice spacing and pion mass; and (4) the lattice renormalization was not performed. We find the current situation not totally satisfactory because although the lattice calculation in the $\Delta I=2$ channel is technically simpler, there is no existing HPV experiment to our knowledge that depends only on the $\Delta I=2$ couplings (see, e.g. Ref.~\cite{Haxton:2013aca} for a summary) so that its comparison with experiments will not be straightforward. In contrast, a successful calculation of $\Delta I=1$ HPV can be directly confronted to the NPDGamma result. Therefore, despite all the technical difficulties, we believe a renewed lattice study of $h_\pi^1$ is extremely worthwhile, and in this work we discuss how the proper application of a chiral EFT in the continuum space may help in alleviating part, if not all, of such difficulties. 

The contents of this paper are as follows. We first introduce the theoretical basis of the $\Delta I=1$ HPV, including the underlying four-quark operators, their Wilson coefficients
and the rigorous definition of the coupling $h_\pi^1$ as a soft-pion matrix element. Next, we review the soft-pion theorem derived in our previous work~\cite{Feng:2017iqb} and present some of the technical details not included in that Letter. Then, we begin the analysis of contraction diagrams by rigorously defining them in terms of three-point correlation functions. With the aid of the partially-quenched chiral perturbation theory~(PQChPT), we derive the theoretical expression for each contraction diagram that contributes to $h_\pi^1$ as a function of the pion mass; such expressions are useful in performing chiral extrapolations from unphysical light quark masses to the physical ones. We point out that there are only a small number of LECs needed to fix the matrix elements, and provide approximate relations between different LECs that may facilitate their global fit. Finally, we briefly discuss the four-quark operators with strange quark fields and draw our conclusions.

\section{Theoretical basis}

We start by reviewing the electroweak interaction Lagrangian of the first two generations of quarks in the SM,
\begin{equation}
\mathcal{L}_{\mathrm{EW}}^q=-eJ_{\rm em}^\mu-\frac{g}{2\sqrt{2}}\left\{W_\mu^+ J_W^\mu+W_\mu^-J_W^{\mu\dagger}\right\}-\frac{g}{2\cos\theta_W}Z_\mu J_Z^\mu ,
\end{equation}
where the electromagnetic, charged weak and neutral weak currents are defined as
\begin{equation}
J_{\rm em}^\mu=\bar{\psi}\gamma^\mu\mathfrak{Q}\psi,\:\:\:J_W^\mu=\bar{\psi}\gamma^\mu(1-\gamma_5)C_+\psi,\:\:\:J_Z^\mu=\frac{1}{2}\bar{\psi}\gamma^\mu(1-\gamma_5)C_3\psi-2\sin^2\theta_WJ_{\rm em}^\mu.
\label{eq:EWcurrents}
\end{equation}
Here, $\psi=(c\:\:u\:\:d\:\:s)^T$ is the quark fields while the matrices $\{\mathfrak{Q},C_+,C_3\}$ are defined as
\begin{equation}
\mathfrak{Q}= \begin{pmatrix}
\frac{2}{3}I_{2} & 0\\
0 & -\frac{1}{3}I_{2}
\end{pmatrix},\:\:\:
C_{+}=\left(\begin{array}{cccc}
0 & 0 & -\sin\theta_{C} & \cos\theta_{C}\\
0 & 0 & \cos\theta_{C} & \sin\theta_{C}\\
0 & 0 & 0 & 0\\
0 & 0 & 0 & 0
\end{array}\right),\:\:\:
C_{3}=\left(\begin{array}{cc}
I_{2} & 0\\
0 & -I_{2}
\end{array}\right),\label{eq:EWmatrices}
\end{equation}
where $\theta_C$ is the Cabibbo angle. A single exchange of a $W$ or $Z$ boson leads to a P-odd interaction between a pair of quarks. At the energy scale $E\ll m_W,m_Z$, the $W$ or $Z$ propagator shrinks to a point, so we obtain effective four-quark interactions involving the product of two weak currents. 

In this work we focus on the $\Delta I=1$ P-violation in nucleon-nucleon interactions, and one may deduce from Eqs.~\eqref{eq:EWcurrents} and \eqref{eq:EWmatrices} that they are dominated by neutral current interactions. An easy way to understand this is to realize that in the $\theta_C\rightarrow 0$ limit, the first and the second generations of quarks completely decouple in the current level, and the charged weak current involving light quarks then reads $J_W^\mu=\bar{u}\gamma^\mu(1-\gamma_5)d$ which is purely an isovector. Therefore, the symmetric combination $J_W^{\mu\dagger}J_{W,\mu}$ can only form $\Delta I=0,2$ objects but not $\Delta I=1$. In reality, the Cabibbo angle is not zero but the charged weak current contribution is suppressed by $\sin^2\theta_C\approx 0.05$ so the neutral current contribution is still dominant. This is an important observation as it identifies the $\Delta I=1$ HWI as one of the very few direct experimental probes of the quark-quark neutral current effects at low energy.

Perturbative QCD modifies the structure of quark-quark weak interactions and introduces operators that do not appear in the original current-current product. Such an effect can be implemented by the QCD renormalization group~(RG) running of the Wilson coefficients of the four-quark operators from the EW scale to the hadronic scale. At low energy, the $\Delta I=1$ HPV can be described by the following Lagrangian~\cite{Kaplan:1992vj}:
\begin{equation}
\mathcal{L}_\mathrm{PV}^w=-\frac{G_F}{\sqrt{2}}\frac{\sin^2\theta_W}{3}\sum_i\left(C_i^{(1)}\theta_i+S_i^{(1)}\theta_i^{(s)}\right),
\label{eq:LPV4quark}
\end{equation}
where\footnote{Notice that Ref.~\cite{Kaplan:1992vj} defines one more operator $\theta_4=\bar{q}_a\gamma^\mu\gamma_5 q_b\bar{q}_b\gamma_\mu\tau_3q_a$, but it is not independent from the rest as  $\theta_4=\theta_1-\theta_2+\theta_3$~\cite{Tiburzi:2012hx}.} 
\begin{align}
\theta_1&= \bar{q}_a\gamma^\mu q_a\bar{q}_b\gamma_\mu\gamma_5\tau_3q_b, &\theta_2&=\bar{q}_a\gamma^\mu q_b\bar{q}_b\gamma_\mu\gamma_5\tau_3q_a ,\nonumber\\
\theta_3&=\bar{q}_a\gamma^\mu\gamma_5 q_a\bar{q}_b\gamma_\mu\tau_3q_b ,\nonumber\\
\theta_1^{(s)}&=\bar{s}_a\gamma^\mu s_a\bar{q}_b\gamma_\mu\gamma_5\tau_3q_b,
&\theta_2^{(s)}&=\bar{s}_a\gamma^\mu s_b\bar{q}_b\gamma_\mu\gamma_5\tau_3q_a ,\nonumber\\
\theta_3^{(s)}&=\bar{s}_a\gamma^\mu\gamma_5 s_a\bar{q}_b\gamma_\mu\tau_3q_b,
&\theta_4^{(s)}&=\bar{s}_a\gamma^\mu\gamma_5 s_b\bar{q}_b\gamma_\mu\tau_3q_a. \label{eq:4qodd}
\end{align}
Here $q=(u\:\:d)^T$ denotes the SU(2) up and down quark fields, and $a,b$ are the color indices. The running of the Wilson coefficients $\{C_i^{(1)},S_i^{(1)}\}$ has been calculated to leading order (LO) in Refs.~\cite{Dai:1991bx,Kaplan:1992vj} and to next-to-leading order (NLO) in Ref.~\cite{Tiburzi:2012hx}. We quote the results of the latter at the scale $\Lambda_\chi\approx1$ GeV:
\begin{eqnarray}
C^{(1)}(\Lambda_{\chi}) & = & \left(\begin{array}{ccc}
-0.055 & 0.810 & -0.627 \end{array}\right),\nonumber \\
S^{(1)}(\Lambda_{\chi}) & = & \left(\begin{array}{cccc}
5.09 & -2.55 & 4.51 & -3.36\end{array}\right).\label{eq:Wilson}
\end{eqnarray}

At the energy scale below $\Lambda_\chi$, the effective degrees of freedom (DOFs) switch from quarks to hadrons. One may then proceed to write down all possible P-odd operators involving the lightest hadronic DOFs following the spirit of EFT. The longest-range P-odd nuclear potential always consists of the pion exchange which, according to the Barton's theorem~\cite{Barton:1961eg}, only survives in the $\Delta I=1$ channel. Thus, the same Lagrangian $\mathcal{L}_\mathrm{PV}^w$ can be expressed at low energy as 
\begin{equation}
\mathcal{L}_\mathrm{PV}^w=-\frac{h_\pi^1}{\sqrt{2}}\bar{N}\left(\vec{\tau}\times\vec{\pi}\right)^3N+...=ih_\pi^1\left(\bar{n}p\pi^--\bar{p}n\pi^+\right)+... \, ,\label{eq:LPVeff}
\end{equation}
where $N=(p\:\:n)^T$ is nucleon isospin doublet and the ellipses denote the remaining HPV interactions of shorter range. Eq.~\eqref{eq:LPVeff} may serve as a definition of the P-odd pion-nucleon coupling constant $h_\pi^1$, but we could equivalently express the latter in terms of a soft-pion matrix element {\color{black}of the P-odd Lagrangian at the origin},
\begin{equation}
h_\pi^1=-\frac{i}{2m_N}\lim_{p_\pi\rightarrow 0}\bigl\langle n\pi^+\bigr|\mathcal{L}_\mathrm{PV}^w(0)\bigl|p\bigr\rangle,
\label{eq:hpidef}
\end{equation}
where $m_N$ is the averaged nucleon mass. The relation above can be obtained by taking the $\left\langle n\pi^+\right|...\left|p\right\rangle$ matrix element at both sides of Eq.~\eqref{eq:LPVeff} and approximating the nucleon spinor product by $\bar{u}_nu_p\approx 2m_N$, neglecting the small neutron-proton mass splitting and any small momentum transfer. Eq.~\eqref{eq:hpidef} serves as the starting point for any first-principle or model-based calculation of $h_\pi^1$. 

\section{From P-odd to P-even matrix element\label{sec:hpimatch}}

This section mainly serves as a review of the results in our previous work~\cite{Feng:2017iqb} with some more technical details added.

\subsection{PCAC relation}

The matrix element in Eq.~\eqref{eq:hpidef}  involves a soft pion in the final state that greatly complicates its analysis. We shall illustrate this point by considering a possible lattice QCD calculation of such a matrix element. First, one needs to choose a form of the interpolator for the $n\pi^+$ state. The most natural choice with the largest overlap with the physical state is obviously a five-quark interpolator, for example, $\varepsilon^{abc}d^a(u^{bT}C\gamma_5d^c)\bar{d}^e\gamma_5 u^e$. Such a choice will however lead to many contraction diagrams, some of which involving up quark propagators between $n$ and $\pi^+$ are noisy and expensive. Another possible choice is a three-quark interpolator with negative parity, as adopted in Ref.~\cite{Wasem:2011zz}, $\varepsilon^{abc}\gamma_5 u^a(d^{bT}C\gamma_5 u^c)$. This choice avoids the calculations of the $n\pi^+$ contraction diagrams, but cannot avoid a large overlap with single-nucleon excited states, e.g., the $N(1535)$. Thus, the use of a three-quark interpolator would be unjustified without properly taking into account the excited-state contaminations. Next, the rescattering effect between the final-state $n\pi^+$ modifies the finite-volume correction on lattice from an exponentially-suppressed effect to a power-suppressed effect. Finally, while we want the final-state pion to have a vanishing momentum squared, lattice QCD only computes matrix elements of on-shell states. 
As a result, the lattice calculation returns not just $h_\pi^1$ but its linear combination with the LECs of total-derivative operators  that must be introduced to compensate the energy difference between the initial $p$ and the final $n\pi^+$~\cite{Beane:2002ca}. Although the leading effect can be canceled by considering the difference between ${p\rightarrow n\pi^+}$ and ${n\rightarrow p\pi^-}$, but the $m_q$-suppressed terms still retain, and they are in principle indistinguishable from the $m_q$-dependent terms of $h_\pi^1$ and lead to a sizable systematic error. 

The situation above can be greatly improved by a simple observation that the pion in the external state plays a special role in QCD: it is the pseudo-Nambu--Goldstone (pNG) boson that arises due to the spontaneously-broken chiral symmetry $\mathrm{SU(2)}_R\times\mathrm{SU(2)}_L\rightarrow \mathrm{SU(2)}_V$. That is, the axial charge operators 
$\hat{Q}_A^i$ do not annihilate the vacuum but produce soft pion states. Consequently, any matrix element involving an external soft pion can be replaced by another matrix element without the soft pion via the partially-conserved axial current~(PCAC) relation,{\color{black}
\begin{equation}
\lim_{p_\pi\rightarrow 0}\bigl\langle a \pi^i\bigr|\hat{O}\bigl|b\bigr\rangle=\frac{i}{F_\pi}\left\langle a \right|[\hat{O},\hat{Q}_A^i]\left|b\right\rangle ,
\label{eq:PCAC}
\end{equation}}
where $F_\pi=92.1$~MeV is the pion decay constant, and $a,b$ are hadrons. This applies exactly to our case: Instead of computing $\left\langle n\pi^+\right|\mathcal{L}_\mathrm{PV}^w\left|p\right\rangle$, one may compute $\left\langle n\right|[\mathcal{L}_\mathrm{PV}^w,\hat{Q}_A^-]\left|p\right\rangle$ which is much simpler. 
To that end, it is beneficial to introduce the following four-quark operators, 
\begin{align}
\theta_1'&= \bar{q}_a\gamma^\mu q_a\bar{q}_b\gamma_\mu\tau_3q_b,
& \theta_2' &=\bar{q}_a\gamma^\mu q_b\bar{q}_b\gamma_\mu\tau_3q_a ,\nonumber\\
\theta_3'&= \bar{q}_a\gamma^\mu\gamma_5 q_a\bar{q}_b\gamma_\mu\gamma_5\tau_3q_b ,\nonumber\\
\theta_1^{(s)\prime}&=\bar{s}_a\gamma^\mu s_a\bar{q}_b\gamma_\mu\tau_3q_b , &\theta_2^{(s)\prime}&=\bar{s}_a\gamma^\mu s_b\bar{q}_b\gamma_\mu\tau_3q_a ,\nonumber\\
\theta_3^{(s)\prime}&=\bar{s}_a\gamma^\mu\gamma_5 s_a\bar{q}_b\gamma_\mu\gamma_5\tau_3q_b,
&\theta_4^{(s)\prime}&=\bar{s}_a\gamma^\mu\gamma_5 s_b\bar{q}_b\gamma_\mu\gamma_5\tau_3q_a. \label{eq:4qeven}
\end{align}

Explicitly evaluating the commutators in the PCAC relation yields
\begin{equation}
\left[\theta_q,\hat{Q}^i_A\right]=i\varepsilon^{3ij}\theta_{q(j)}',\qquad
\left[\theta_q^{(s)},\hat{Q}^i_A\right]=i\varepsilon^{3ij}\theta_{q(j)}^{(s)'},
\end{equation}
where $\{\theta_{q(j)}',\theta_{q(j)}^{(s)'}\}$ are defined as the P-even four-quark operators $\{\theta_{q}',\theta_{q}^{(s)'}\}$ in Eq.~\eqref{eq:4qeven} with the replacement $\tau_3\rightarrow \tau_j$. 
Thus, the P-odd hadronic matrix elements $\left\langle n\pi^+\right|\theta_q^{}\left|p\right\rangle$ and $\left\langle n\pi^+\right|\theta_q^{(s)}\left|p\right\rangle$ can be mapped to the P-even hadronic matrix elements $\left\langle n\right|\theta_{q(j)}'\left|p\right\rangle$ and $\left\langle n\right|\theta_{q(j)}^{(s)'}\left|p\right\rangle$ by PCAC, respectively. We may go one step further by transforming the latter into flavor-diagonal hadronic matrix elements through isospin rotation. By doing so, the operators in the matrix elements turn into those in Eq. \eqref{eq:4qeven}. The final result is
\begin{equation}
\lim_{p_\pi\rightarrow 0}\bigl\langle n\pi^+\bigr|\mathcal{L}_{\mathrm{PV}}^{w}(0)\bigl|p\bigr\rangle\approx-\frac{\sqrt{2}i}{F_\pi}\left\langle p\right|\mathcal{L}_{\mathrm{PC}}^{w}(0)\left|p\right\rangle=\frac{\sqrt{2}i}{F_\pi}\left\langle n\right|\mathcal{L}_{\mathrm{PC}}^{w}(0)\left|n\right\rangle,\label{eq:PVtoPC}
\end{equation}
where $\mathcal{L}_\mathrm{PC}^w$ is an auxiliary P-even Lagrangian,
\begin{equation}
\mathcal{L}_\mathrm{PC}^w=-\frac{G_F}{\sqrt{2}}\frac{\sin^2\theta_W}{3}\sum_i\left(C_i^{(1)}\theta_i'+S_i^{(1)}\theta_i^{(s)\prime}\right).
\end{equation}
The reader should be alerted that $\mathcal{L}_\mathrm{PC}^w$ is not the actual P-conserving weak four-quark interaction in SM, but simply an auxiliary Lagrangian introduced to facilitate the calculation of $h_\pi^1$, and the Wilson coefficients have to be identical with those in $\mathcal{L}_\mathrm{PV}^w$.

At this level we have successfully mapped a P-odd $N\rightarrow N'\pi$ matrix element to a flavor-diagonal, P-even $N\rightarrow N$ matrix element. The right hand side (RHS) of Eq.~\eqref{eq:PVtoPC} can be rewritten in terms of the neutron-pion mass splitting $(\delta m_N)_{4q}\equiv \left(m_n-m_p\right)_{4q}$ {\color{black}induced by the P-even Lagrangian $\mathcal{L}_\mathrm{PC}^w$,}
\begin{equation}
(\delta m_N)_{4q}=\frac{1}{m_N}\left\langle p\right|\mathcal{L}_{\mathrm{PC}}^{w}(0)\left|p\right\rangle=-\frac{1}{m_N}\left\langle n\right|\mathcal{L}_{\mathrm{PC}}^{w}(0)\left|n\right\rangle.\label{eq:FH_thm}
\end{equation}
Thus, combining Eqs. \eqref{eq:hpidef}, \eqref{eq:PVtoPC} and \eqref{eq:FH_thm}, we obtain an approximate relation between $h_\pi^1$ and $(\delta m_N)_{4q}$,
\begin{equation}
F_\pi h_\pi^1\approx-\frac{(\delta m_N)_{4q}}{\sqrt{2}},\label{eq:central}
\end{equation}
which is one of the central results in Ref.~\cite{Feng:2017iqb}.

We would like to point out that the idea above is not at all new. To our knowledge, the first application of PCAC in the study of the $\Delta S=0$ weak pion-baryon vertex appeared in Ref.~\cite{Fischbach:1968zz} in the late 60s; it was also adopted in the DDH paper~\cite{Desplanques:1979hn} as well as Ref.~\cite{Kaiser:1988bt} as a starting point of their model-based estimation of $h_\pi^1$. The originality of Ref.~\cite{Feng:2017iqb} is really not in its application of PCAC, but rather in its quantitative analysis of the higher-order corrections which determines the degree of accuracy of the PCAC result, as we shall describe later.

\subsection{Chiral perturbation theory analysis}

The PCAC relation in Eq.~\eqref{eq:PCAC} holds rigorously only in the exact chiral limit, i.e., when $m_\pi=0$. For example, it predicts that the matrix element at the left hand side should vanish if $\hat{O}$ is chirally-invariant, which is obviously incorrect. Here we shall provide an immediate counter-example in a closely-related problem, namely the study of the P, T-odd pion-nucleon coupling $\bar{g}_\pi^i$ induced by higher-dimensional operators. Eq.~\eqref{eq:PCAC} suggests that chirally-invariant operators such as the Weinberg three-gluon operator $f^{ABC}\tilde{G}^A_{\mu\nu}G^{B\nu}_{\rho}G^{C\rho\mu}$ would not contribute to $\bar{g}_\pi^i$; however, we know in reality that the contribution of such an operator is non-zero, but just suppressed by powers of $m_\pi$~\cite{Engel:2013lsa}. Therefore, a truly practical application of the PCAC relation will need to take into account all the $m_\pi$-related corrections to the level of desired precision.

The above-mentioned task is made possible by recasting the PCAC statement in the language of chiral perturbation theory (ChPT), where Eq.~\eqref{eq:PCAC} then becomes a simple consequence of two observables sharing the same LEC at the tree level. Higher-order corrections such as loop diagrams and counterterms to the left and the right sides of the equation can be computed order-by-order; any mismatch will then signifies a quantifiable violation of the tree-level matching. This idea was born of the in-depth studies of the P, T-odd pion-nucleon coupling $\bar{g}_\pi^i$~\cite{deVries:2012ab,Mereghetti:2015rra,deVries:2015una,Seng:2016pfd,deVries:2016jox,Cirigliano:2016yhc}, and Ref.~\cite{Feng:2017iqb} constitutes its first implementation in HPV. Below we shall describe the method in detail.

Let us start by introducing the basic ingredients in a two-flavor ChPT with nucleons and pions. For that purpose it is instructive to first look at the QCD Lagrangian with two quark flavors,
\begin{equation}
\mathcal{L}_\mathrm{QCD}=\bar{q}_Li\slashchar{D}q_L+\bar{q}_Ri\slashchar{D}q_R-\bar{q}_RM_qq_L-\bar{q}_LM_q^\dagger q_R-\frac{1}{4}G^a_{\mu\nu}G^{a\mu\nu}, \label{eq:LQCD}
\end{equation}
where $q_{R,L}=(1/2)(1\pm \gamma_5)q$ is the right/left-handed component of the quark field. The quark mass matrix is given by $M_q=\mathrm{diag}(m_u, m_d)$. Direct inspection of Eq.~\eqref{eq:LQCD} shows that the Lagrangian is invariant under the following $\mathrm{SU(2)}_R\times\mathrm{SU(2)}_L$ chiral rotation,
\begin{equation}
q_R\rightarrow Rq_R, \:\:\:q_L\rightarrow Lq_L ,
\end{equation}
in the limit of vanishing quark masses. Furthermore, if one would assume a transformation rule $M_q\rightarrow RM_qL^\dagger$ of the quark mass matrix, then $\mathcal{L}_\mathrm{QCD}$ would be chirally-invariant even with the existence of the quark masses. This is the so-called spurion trick to take into account symmetry breaking terms.

ChPT involves writing down all possible operators with hadronic DOFs that are consistent with the symmetry of $\mathcal{L}_\mathrm{QCD}$ under chiral rotation. There are infinitely many terms of such a kind. Thus, they have to be arranged according to a power counting scheme such that in any given order there are only a finite number of terms. The pions are contained in the matrix $U$ defined as
\begin{equation}
U=\exp\left\{\frac{i\vec{\pi}\cdot\vec{\tau}}{F_0}\right\} ,
\end{equation}
where $F_0$ is the pion decay constant in the chiral limit. It transforms as $U\rightarrow RUL^\dagger$ under the chiral rotation. The chiral Lagrangian of pions at LO consists of only two terms,
\begin{equation}
\mathcal{L}_\pi=\frac{F_0^2}{4}\mathrm{Tr}\left[\partial_\mu U\partial^\mu U^\dagger\right]+\frac{F_0^2B_0}{2}\mathrm{Tr}\left[M_qU^\dagger+UM_q^\dagger\right].
\end{equation}
In particular, the second term gives rise to the pion mass at LO, $m_\pi^2=B_0(m_u+m_d)$.

In the baryon sector, the nucleon doublet $N$ appears as a matter field and can be chosen to transform as $N\rightarrow KN$ under the chiral rotation, where $K$ is a spacetime-dependent matrix defined through the transformation property of $u=\sqrt{U}$,
\begin{equation}
u\rightarrow RuK^\dagger=KuL^\dagger.
\end{equation}  
The chiral Lagrangian of nucleon at LO reads
\begin{equation}
\mathcal{L}_N=\bar{N}\left(i\slashchar{\mathcal{D}}-m_0\right)N+\frac{g_0}{2}\bar{N}\gamma^\mu\gamma_5u_\mu N ,
\label{eq:LNrel}\end{equation}
where $m_0$ and $g_0$ are the nucleon mass and axial coupling constant in the chiral limit, the chiral covariant derivative is defined as $\mathcal{D}_\mu=\partial_\mu+\Gamma_\mu$, and 
\begin{equation}
 \Gamma_\mu = \frac12 \left(u^\dagger \partial_\mu u+u\partial_\mu u^\dagger\right), \qquad 
 u_\mu=i\left(u^\dagger\partial_\mu u-u\partial_\mu u^\dagger\right)
 \label{eq:ugamma}
\end{equation}
are the vector connection and axial vector, respectively, in the absence of external fields. 

The na\"{\i}ve application of the nucleon Lagrangian in Eq.~\eqref{eq:LNrel} will cause problems as it contains a large bare nucleon mass term that needs to be treated as a large energy scale. There are different schemes introduced to tackle this issue, for example the heavy baryon~(HB) expansion~\cite{Jenkins:1990jv,Bernard:1992qa,Bernard:1995dp}, the infrared regularization~\cite{Ellis:1997kc,Becher:1999he} and the extended on-mass-shall scheme (EOMS)~\cite{Gegelia:1999gf,Fuchs:2003qc}. While the first approach is technically simplest, the other two approaches possess extra advantages in preserving the analytic structures of the amplitude. In this work we are only interested in static matrix elements that are insensitive to the analytic behaviors around the threshold, so we shall just adopt the simplest HB approach. Below we shall briefly summarize its most important results, and interested readers may refer to standard textbooks such as Ref.~\cite{Scherer:2012xha} for details. In this approach, a redefinition of the nucleon field is performed to remove the large bare mass term in the Lagrangian, Eq.~\eqref{eq:LNrel}. As a consequence, the nucleon field $N$ is effectively replaced by its ``light" component $N_v$ which appears as a massless excitation, and the Dirac structures are effectively reduced as $\gamma^\mu\rightarrow v^\mu$ and $\gamma^\mu\gamma_5\rightarrow 2S^\mu$ where $v^\mu$ is a constant four-velocity vector and $S^\mu=i\gamma_5\sigma^{\mu\nu}v_\nu/2$ is the nucleon spin-matrix satisfying $S\cdot v=0$. With this, the LO nucleon Lagrangian becomes
\begin{equation}
\mathcal{L}_N\rightarrow \bar{N}_viv\cdot\mathcal{D}N_v+g_0\bar{N}_vu_\mu S^\mu N_v,
\end{equation}
and the bare nucleon propagator reads $i/(v\cdot k+i\epsilon)$ where $k^\mu$ is the residual $\mathcal{O}(p)$ momentum of the nucleon which is related to the full nucleon momentum $p_N^\mu$ by $p_N^\mu=m_Nv^\mu+k^\mu$.

The effects of the $\Delta I=1$ four-quark operators can be most easily implemented to the chiral Lagrangian by adding the P-odd and P-even Lagrangian to obtain
\begin{equation}
\mathcal{L}_\mathrm{tot}^w=\mathcal{L}_\mathrm{PV}^w+\mathcal{L}_\mathrm{PC}^w=-\frac{G_F}{\sqrt{2}}\frac{\sin^2\theta_W}{3}\sum_i\left(C_i^{(1)}\tilde{\theta}_i+S_i^{(1)}\tilde{\theta}_i^{(s)}\right) ,
\end{equation}
where
\begin{align}
\tilde{\theta}_1&=2\bar{q}_a\gamma^\mu q_{a}\bar{q}_{bR}\gamma_\mu\tau_3q_{bR}, &\tilde{\theta}_2&=2\bar{q}_a\gamma^\mu q_b\bar{q}_{bR}\gamma_\mu\tau_3q_{aR}, \nonumber\\
\tilde{\theta}_3&=2\bar{q}_a\gamma^\mu\gamma_5 q_a\bar{q}_{bR}\gamma_\mu\tau_3q_{bR}, \nonumber\\
\tilde{\theta}_1^{(s)}&=2\bar{s}_a\gamma^\mu s_a\bar{q}_{bR}\gamma_\mu\tau_3q_{bR},
&\tilde{\theta}_2^{(s)}&=2\bar{s}_a\gamma^\mu s_b\bar{q}_{bR}\gamma_\mu\tau_3q_{aR},\nonumber\\
\tilde{\theta}_3^{(s)}&=2\bar{s}_a\gamma^\mu\gamma_5 s_a\bar{q}_{bR}\gamma_\mu\tau_3q_{bR},
&\tilde{\theta}_4^{(s)}&=2\bar{s}_a\gamma^\mu\gamma_5 s_b\bar{q}_{bR}\gamma_\mu\tau_3q_{aR}.
\end{align} 
One immediately observes that all the operators listed above would be invariant under SU(2) chiral rotation if the matrix $\tau_3$ would transform as $\tau_3\rightarrow R\tau_3 R^\dagger$. Therefore, the effect of the combined operators can be implemented to the chiral Lagrangian through a single Hermitian, traceless spurion $X_R=u^\dagger \tau_3 u$ that transforms as $X_R\rightarrow KX_RK^\dagger$. There is only one available operator at LO,
\begin{equation}
\mathcal{L}_{\mathrm{tot,LO}}^w=\alpha\bar{N}_vX_RN_v=\alpha\bar{N}_v\tau_3N_v-\frac{\sqrt{2}i}{F_0}\alpha\left(
\bar{n}_vp_v\pi^--\bar{p}_vn_v\pi^+\right)+... ,
\end{equation}
where the quantity $\alpha$ is an unknown LEC that describes the strength of the dressed weak interaction. Upon expanding the Lagrangian with respect to the pion fields, we find that the first and the second terms correspond exactly to the neutron-proton mass splitting and the P-odd pion-nucleon coupling, respectively. Since they share the same unknown coefficient $\alpha$, we immediately obtain Eq.~\eqref{eq:central} which is the PCAC prediction.

\begin{figure}[t]
	\includegraphics[scale=0.15]{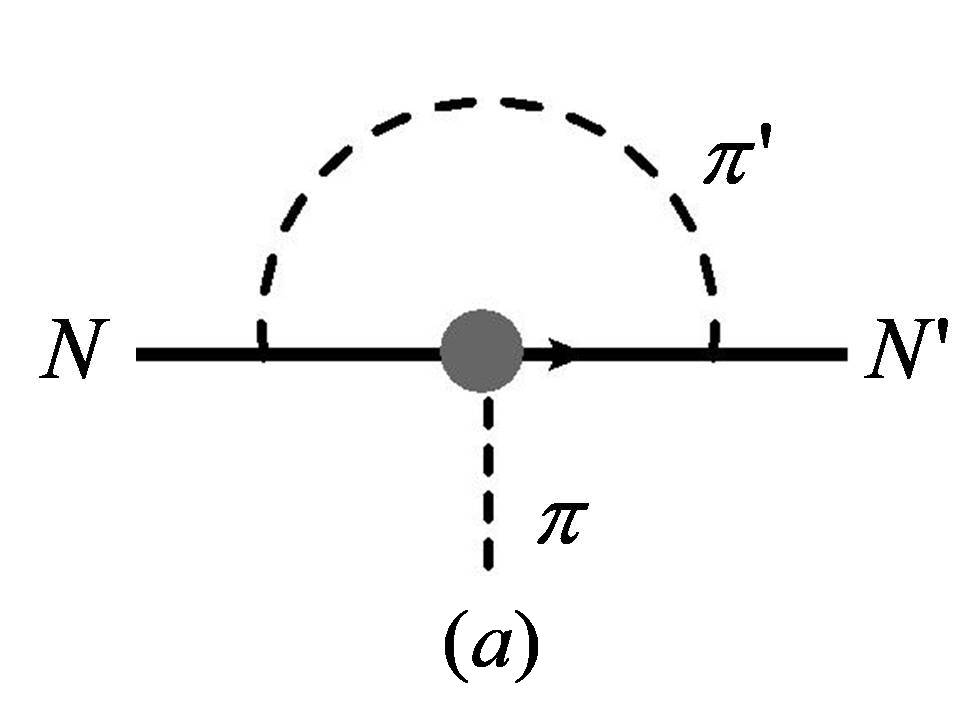}
	\includegraphics[scale=0.15]{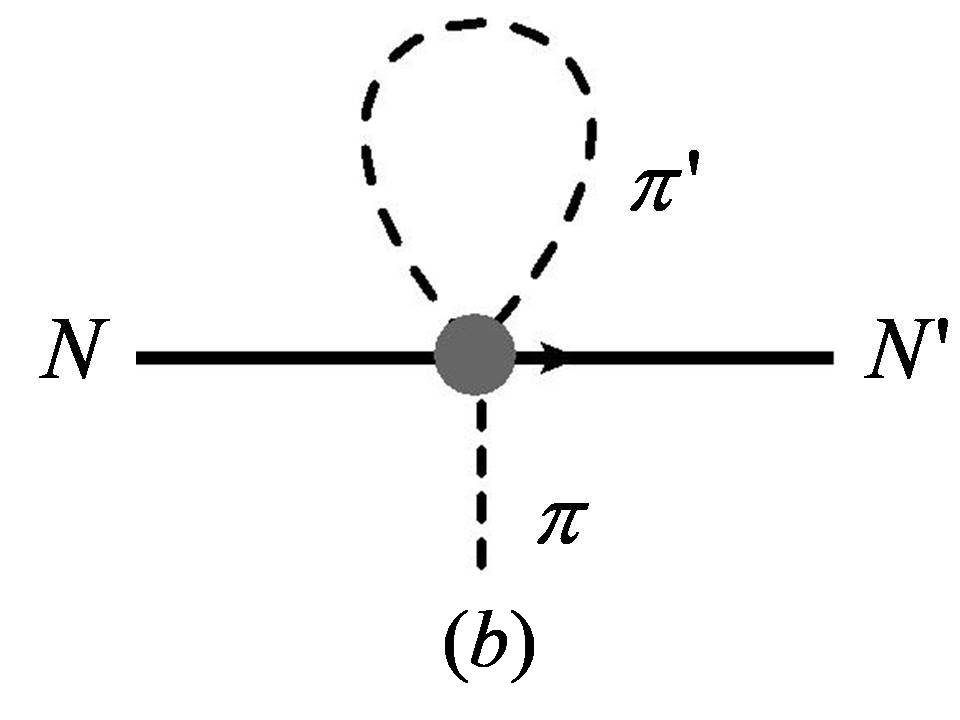}
	\includegraphics[scale=0.15]{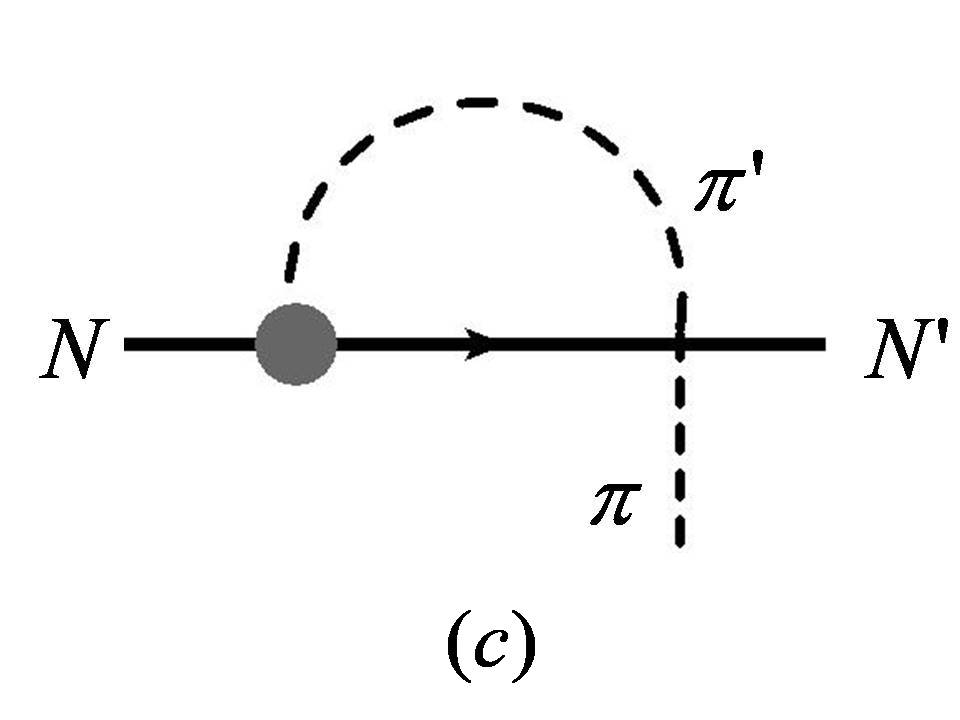}
	\includegraphics[scale=0.15]{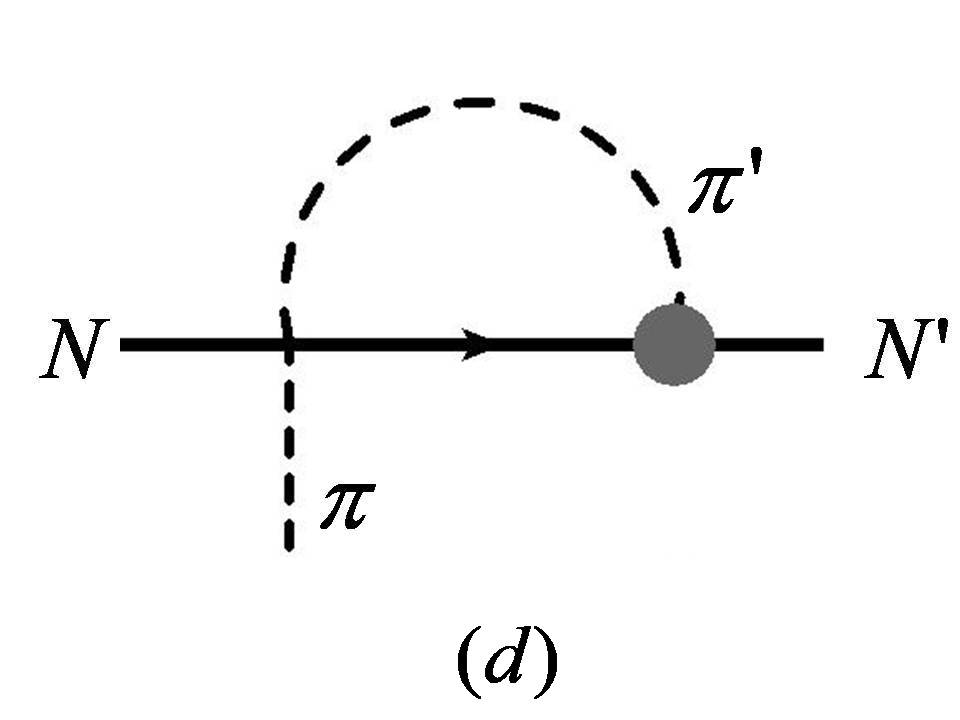}
	\includegraphics[scale=0.15]{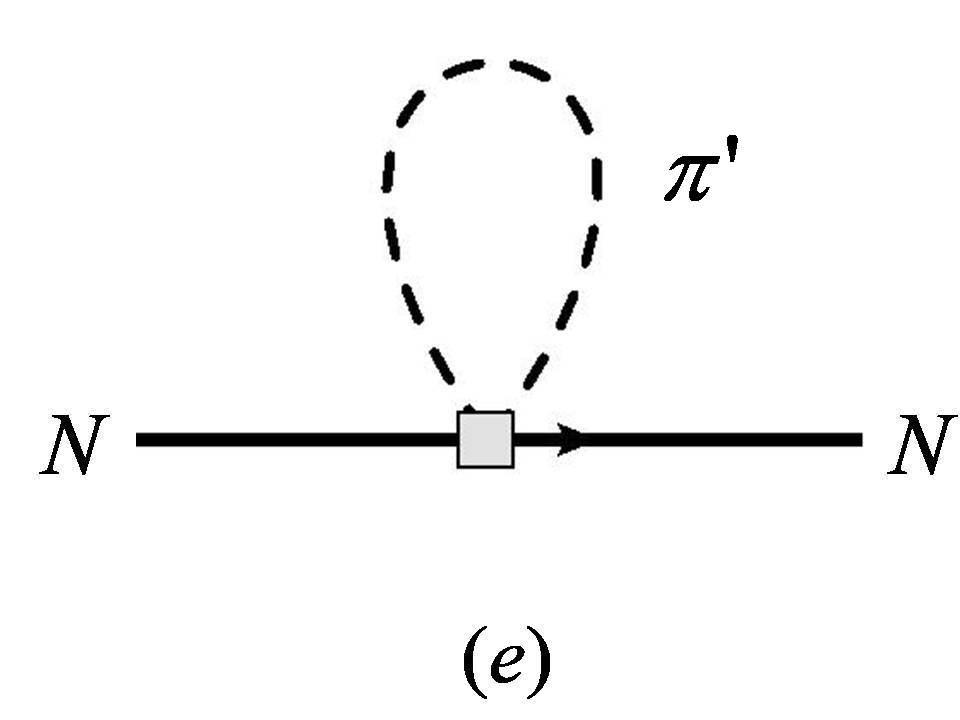}
	\includegraphics[scale=0.15]{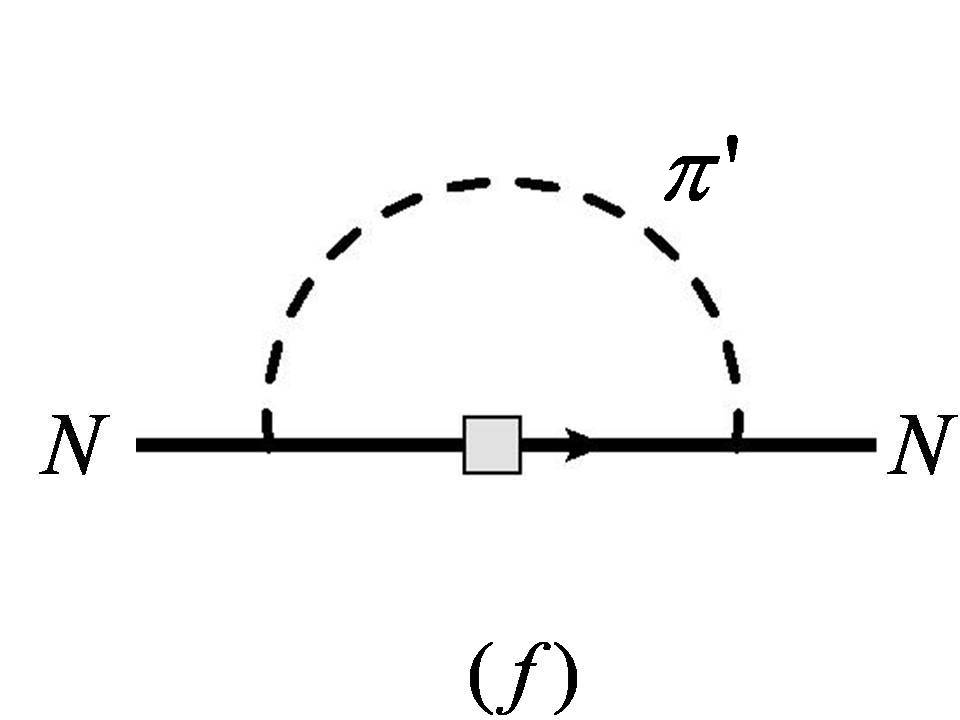}
	\caption{\label{fig:hpimNloop}1PI diagrams involved in the analysis of $h_\pi^1$-$(\delta m_N)_{4q}$ matching relation at one loop.}
\end{figure}

Our conclusion above is based on a tree-level analysis of ChPT, so it is legitimate to ask whether it still holds under higher-order corrections, including both the long-range (one loop) and short-range (counterterm) corrections. The way to proceed is to compute explicitly the higher-order corrections to both the left and the right sides of Eq.~\eqref{eq:central}. First, the one-particle-irreducible (1PI) diagram corrections to $h_\pi^1$ and $(\delta m_N)_{4q}$ at one loop are depicted by the first four diagrams and the last two diagrams in Fig.~\ref{fig:hpimNloop}, respectively. They give
\begin{eqnarray}
\delta\left( h_\pi^1\right)_\mathrm{1PI}&=&\left(\frac{g_0^2}{F_\pi^2}I_a-\frac{5}{6F_\pi^2}I_e\right)h_\pi^1, \nonumber\\
\delta\left((\delta m_N)_{4q}\right)_\mathrm{1PI}&=&\left(\frac{g_0^2}{F_\pi^2}I_a-\frac{1}{F_\pi^2}I_e\right)(\delta m_N)_{4q},
\end{eqnarray}
where $I_a$ and $I_e$ are two standard loop functions (see, e.g., Ref.~\cite{Seng:2016pfd}) that often occur in one-loop calculations with HBChPT,
\begin{eqnarray}
I_a&=&\mu^{4-d}\int\frac{d^dk}{(2\pi)^d}\left(S\cdot k\right)^2\left(\frac{i}{v\cdot k+i\epsilon}\right)^2\frac{i}{k^2-m_\pi^2+i\epsilon} =\frac{3m_\pi^2}{64\pi^2}\left(\lambda-\frac{2}{3}+\ln\frac{\mu^2}{m_\pi^2}\right),\nonumber\\
I_e&=&\mu^{4-d}\int\frac{d^dk}{(2\pi)^d}\frac{i}{k^2-m_\pi^2+i\epsilon}=-\frac{m_\pi^2}{16\pi^2}\left(\lambda+\ln\frac{\mu^2}{m_\pi^2}\right),
\end{eqnarray}
with $\lambda={2}/{(4-d)}-\gamma_E+\ln 4\pi+1$.
Notice that dimensional regularization is used to regularize the ultraviolet (UV) divergences in the loops.

Next, we consider the corrections from the LECs of order $\mathcal{O}(m_\pi^2/\Lambda_\chi^2)$. There are only two independent operators one could write down at this order,
\begin{equation}
\mathcal{L}^w_{\mathrm{tot,NLO}}=\tilde{c}_1\bar{N}_v\left\{\chi_+,X_R\right\}
N_v+\tilde{c}_2\mathrm{Tr}\left(\chi_+\right)\bar{N}_vX_RN_v,
\end{equation}
and their contributions read
\begin{eqnarray}
\delta(h_\pi^1)_{\mathrm{LEC}}&=& 
-\frac{8\sqrt{2}}{F_\pi}B_0\bar{m}(\tilde{c}_1+\tilde{c}_2)\,,\nonumber\\
\delta((\delta m_N)_{4q})_{\mathrm{LEC}}&=& 
16B_0\bar{m}(\tilde{c}_1+\tilde{c}_2),\label{eq:LwNLO}
\end{eqnarray} 
where $\bar{m}=(m_u+m_d)/2$ is the average light quark mass.\footnote{Notice that no isospin symmetry is assumed here.} Finally, one also needs to include the standard pion~\cite{Gasser:1983yg} and nucleon~\cite{Bernard:1992qa,Bernard:1995dp} wavefunction renormalization as well as the higher-order correction to the pion decay constant $F_\pi$,\footnote{We take this opportunity to correct a typo in Eq. (13) of Ref. \cite{Feng:2017iqb}: The denominator in the first term of $\delta (F_\pi)$ should be $F_\pi$ instead of $F_\pi^2$.}
\begin{eqnarray}
\sqrt{Z_\pi}-1&=&\frac{1}{3F_\pi^2}I_e-\frac{m_\pi^2}{F_\pi^2}l_4 ,\nonumber\\
Z_N-1&=&\frac{3g_0^2}{F_\pi^2}I_a-\frac{m_\pi^2}{2\pi^2F_\pi^2}B_{20} ,\nonumber\\
\delta\left(F_\pi\right)&=&-\frac{1}{F_\pi}I_e+\frac{m_\pi^2}{F_\pi}l_4 , \label{eq:ZN}
\end{eqnarray}
where $l_4$ and $B_{20}$ are LECs in the mesonic ChPT and HB ChPT introduced in Ref.~\cite{Gasser:1983yg} and Ref.~\cite{Ecker:1994pi}, respectively.

After grouping everything together one obtains~\cite{Feng:2017iqb}
\begin{eqnarray}
\delta\left(F_\pi h_\pi^1\right)&=&\left(\frac{4g_0^2}{F_\pi^2}I_a-\frac{1}{F_\pi^2}I_e-\frac{m_\pi^2}{2\pi^2F_\pi^2}B_{20}\right)F_\pi h_\pi^1-8\sqrt{2}B_0\bar{m}(\tilde{c}_1+\tilde{c}_2),\nonumber\\
\delta\left((\delta m_N)_{4q}\right)&=&\left(\frac{4g_0^2}{F_\pi^2}I_a-\frac{1}{F_\pi^2}I_e-\frac{m_\pi^2}{2\pi^2F_\pi^2}B_{20}\right)(\delta m_N)_{4q}+16B_0\bar{m}(\tilde{c}_1+\tilde{c}_2).\label{eq:deltahpimN}
\end{eqnarray}
Namely, the higher-order corrections to $F_\pi h_\pi^1$ and $(\delta m_N)_{4q}$ (up to $\mathcal{O}(m_\pi^2/\Lambda_\chi^2)$) satisfies exactly the same matching relation in Eq.~\eqref{eq:central}. Hence, the accuracy of such a relation is {\color{black}ideally expected to be} better than $m_\pi^2/\Lambda_\chi^2\sim 1\%$ when the pion mass takes its physical value. {\color{black}Yet, several studies of the pion-nucleon scattering and the nucleon mass corrections suggest that in baryon ChPT the chiral power counting could break down at a lower scale $\Lambda_\chi'\sim 400$ MeV~\cite{Bernard:2003rp,Djukanovic:2006xc,Bernard:2007zu}, but even in that case the accuracy of the relation is still better than $m_\pi^2/\Lambda_\chi^{\prime 2}\sim 10\%$.} This is the most important observation in Ref.~\cite{Feng:2017iqb}. It shows that the PCAC relation does not simply serve for an illustration purpose but is also quantitatively accurate. Hence, for all practical purposes, it is sufficient to calculate $(\delta m_N)_{4q}$ or equivalently $\left\langle p\right|\mathcal{L}_\mathrm{PC}^w\left|p\right\rangle$ on lattice which is much easier than $h_\pi^1$. We shall also contrast the result here with similar studies of $\bar{g}_\pi^i$. For the latter, the degree of accuracy of the tree-level matching depends critically on the involved P, T-odd operators: For the QCD $\theta$-term and the quark chromo-electric dipole moments, the relation is preserved by one-loop corrections but violated by counterterms; for P, T-odd four-quark operators, the relation is violated by both one-loop corrections and counterterms~\cite{Seng:2016pfd,deVries:2016jox}. A reason of such worse behavior is that a P, T-odd source may introduce a linear term in the pion field which has to be rotated away. Such a rotation gives rise to an extra piece in the tree-level matching relation in addition to the PCAC prediction, and the matching of this extra piece is usually not preserved at higher orders.  

\section{\label{sec:contract}Baryon interpolators and the contraction diagram analysis}

The objective of this paper is to investigate the possibility of performing a high-precision calculation of the coupling $h_\pi^1$ on the lattice. To that end, it is instructive to go through the existing limitations of the original Wasem calculation (as pointed out in Ref.~\cite{WalkerLoud}) and ask ourselves how many of them can be properly taken into account from a theoretical point of view. Clearly, our strategy in Ref.~\cite{Feng:2017iqb} avoids the unjustified application of the $N\pi$ interpolator and alleviates the effect of the finite-volume correction, but does not resolve the other issues and therefore is not the end of the story. We shall devote the rest of this paper to the discussion of the remaining problems that could be at least addressed partially in a continuous field theory. In particular, we shall discuss the properties of different Wick contraction diagrams that occur in the the calculation of $(\delta m_N)_{4q}$ on lattice and the chiral extrapolation of the lattice result. We shall assume isospin symmetry throughout the rest of the paper.

We start by reviewing the standard procedure to compute a hadronic matrix element of a four-quark operator on lattice. Consider the following $\Delta I=1$, P-even four-quark operator without strange quark fields:
\begin{equation}
\hat{O}=\bar{q}^a\Gamma^\mu q^a\bar{q}^b\tau_3\Gamma_\mu q^b,
\end{equation}
where $\Gamma^\mu$ is a general Dirac structure. To be concrete we choose the color contraction to be of the type $aabb$ but the discussion of the operators with color contraction of the type $abba$ proceeds exactly the same way. We are interested in the matrix element of $\hat{O}$ with respect to the static proton state. Following the spirit of lattice QCD, we define a two-point and a three-point correlation function (with $\tau>\tau_0>\tau'$): 
\begin{eqnarray}
F_2(\tau,\tau')&=&\left\langle 0\right|\hat{O}_p(\vec{x},\tau)\hat{\bar{O}}_p(\vec{x}',\tau')\left|0\right\rangle ,\nonumber\\
F_3(\tau,\tau_0,\tau')&=&\left\langle 0\right|\hat{O}_p(\vec{x},\tau)\hat{O}(\vec{x}_0,\tau_0)\hat{\bar{O}}_p(\vec{x}',\tau')\left|0\right\rangle ,
\end{eqnarray}
where $\hat{O}_p$ is the so-called proton interpolator which is an operator with the same quantum number as the proton state, $\vec{x},\vec{x}_0,\vec{x}'$ are some fixed spatial points on the lattice and $\tau,\tau_0,\tau'$ are Euclidean time: $\tau=it$. One may insert a compete set of states between any two operators,
\begin{equation}
1 = \sum_n\frac{1}{2E_n L^3}\left|n\right\rangle \left\langle n\right| ,
\end{equation}
with $L$ the lattice size. Here, the summation also runs over all possible quantized momentum modes.\footnote{The three-momentum is quantized as $2\pi \vec{m}/L$, with $\vec m\in\mathbb{Z}^3$ a three-dimensional vector of integers, in a finite volume with periodic boundary conditions.}  When $\tau-\tau_0\rightarrow+\infty$ and $\tau_0-\tau'\rightarrow+\infty$, only the static proton contribution that scales as $\exp\{-m_N(\tau-\tau')\}$ survives. Therefore one obtains
\begin{equation}
\left\langle p(\vec 0\,)\right|\hat{O}(0)\left|p(\vec 0\,)\right\rangle=2m_N L^3\lim_{\substack{\tau-\tau_0\rightarrow+\infty\\ \tau_0-\tau'\rightarrow+\infty}}\frac{ F_3(\tau,\tau_0,\tau')}{F_2(\tau,\tau')}.
\end{equation} 
In actual lattice calculations, one may choose to sum over a part of the spatial points $\{\vec{x},\vec{x}_0,\vec{x}'\}$ in order to improve the signal; this is however not directly relevant to our study of contraction diagrams so we shall not discuss it any further.

Using the Wick theorem, the three-point correlation function $F_3(\tau,\tau_0,\tau')$ can be separated into different terms according to the types of quark contractions. To rigorously define a specific contraction as a mathematical quantity, we need to first choose the three-quark interpolating operator of the ground-state spin-1/2 baryons. Here we find the most convenient choice for theoretical manipulation to be
\begin{equation}
\chi(q_1,q_2,q_3;\vec{x},\tau)=\varepsilon^{abc}\gamma_5\gamma^\mu q_1^a(\vec{x},\tau)q_2^{b\mathrm{T}}(\vec{x},\tau)C\gamma_\mu q_3^c(\vec{x},\tau),\label{eq:interpolator}
\end{equation}
where $C$ is the charge conjugation matrix.
It satisfies the following symmetry relations under the exchange of quark flavors,
\begin{eqnarray}
\chi(q_1,q_3,q_2;\vec{x},\tau)&=&\chi(q_1,q_2,q_3;\vec{x},\tau),\nonumber\\
\chi(q_1,q_2,q_3;\vec{x},\tau)&=&-\chi(q_2,q_1,q_3;\vec{x},\tau)-\chi(q_3,q_2,q_1;\vec{x},\tau), \label{eq:flavorsymmetry}
\end{eqnarray}
which are necessary to single-out the spin-1/2 ground state baryons (notice that this choice of three-quark interpolator is not unique; other choices are discussed in Appendix~\ref{sec:Otherint}). We may define the proton interpolator as: $\hat{O}_p=(1/2)\chi(d,u,u)$ where the factor 1/2 is just an arbitrary choice of normalization. Once this is fixed,  the interpolators of all other spin-1/2 ground state baryons can also be determined by means of quark model~\cite{Donoghue:1985rk}: for instance, $\hat{O}_n=\chi(d,d,u)$, $\hat{O}_{\Sigma^0}=(1/\sqrt{2})\chi(s,d,u)$ and $\hat{O}_\Lambda=(1/\sqrt{6})\left[\chi(u,d,s)-\chi(d,u,s)\right]$.

\begin{figure}[t]
	\begin{centering}
		\includegraphics[scale=0.2]{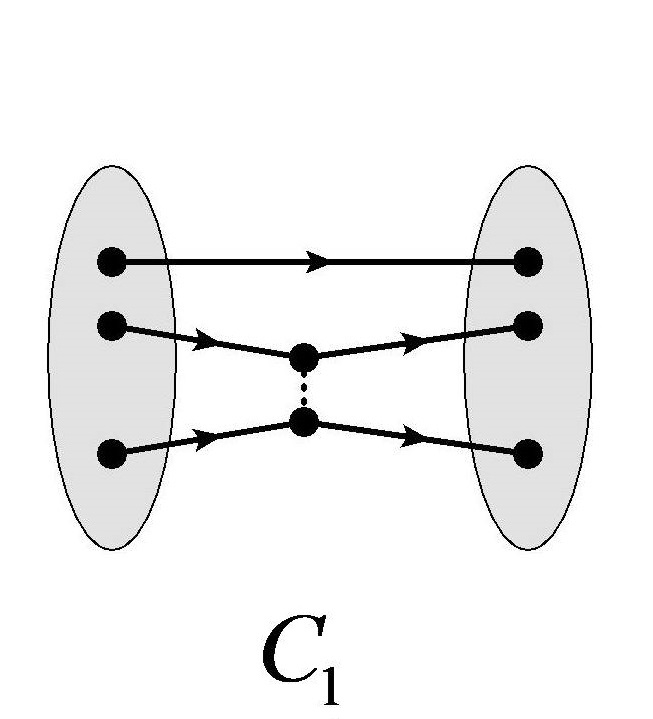} 
		\includegraphics[scale=0.2]{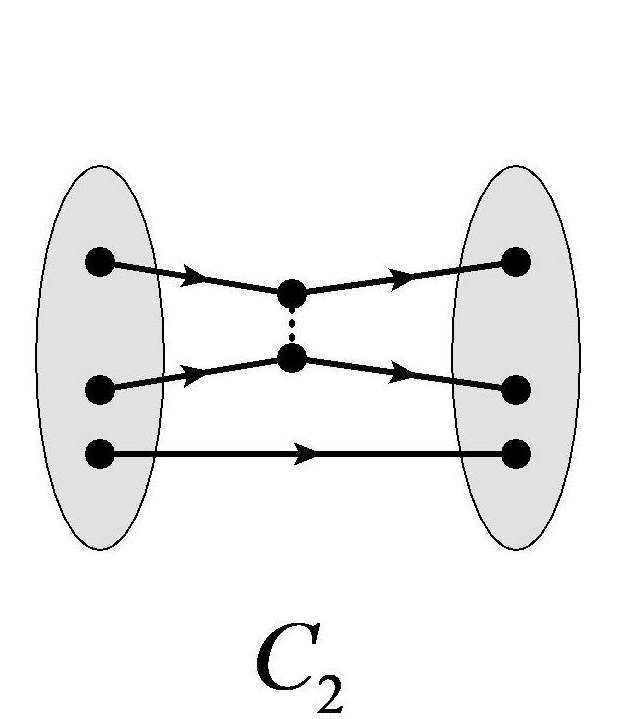}
		\includegraphics[scale=0.2]{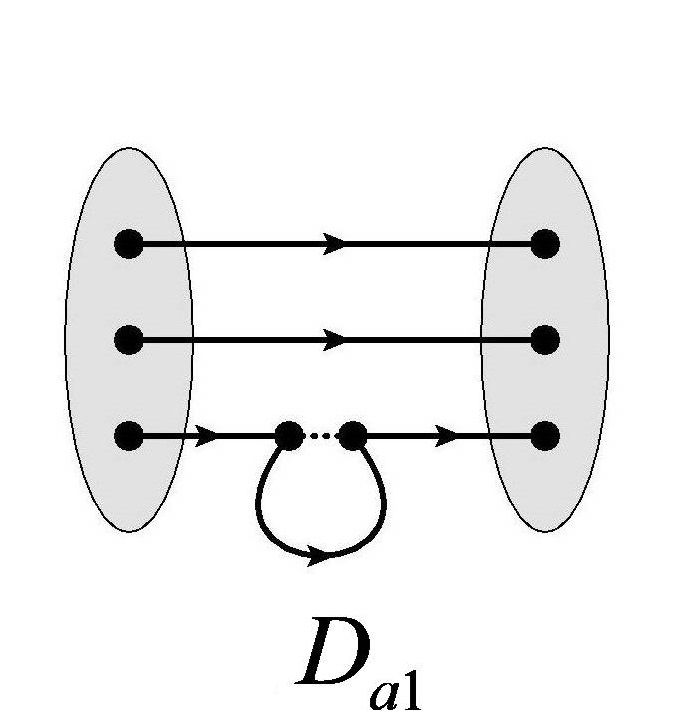}
		\includegraphics[scale=0.2]{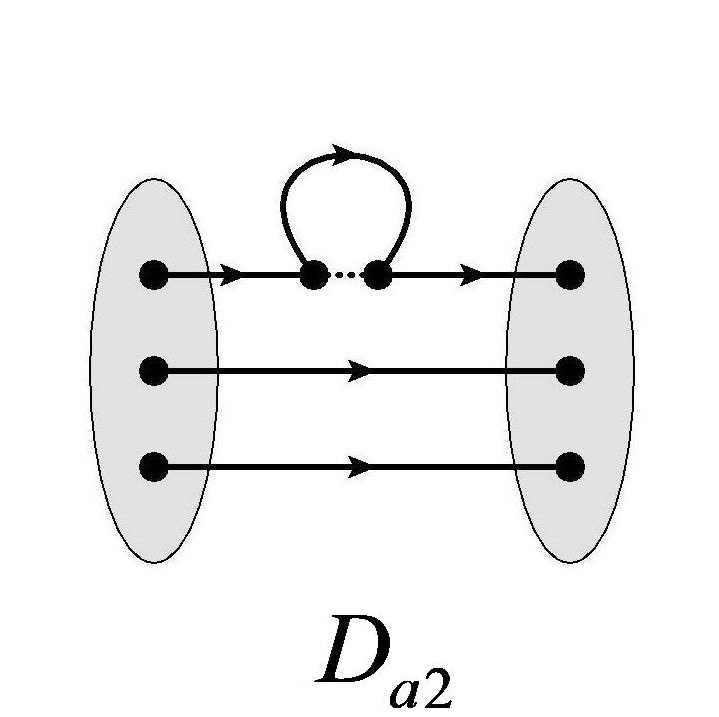}
		\includegraphics[scale=0.2]{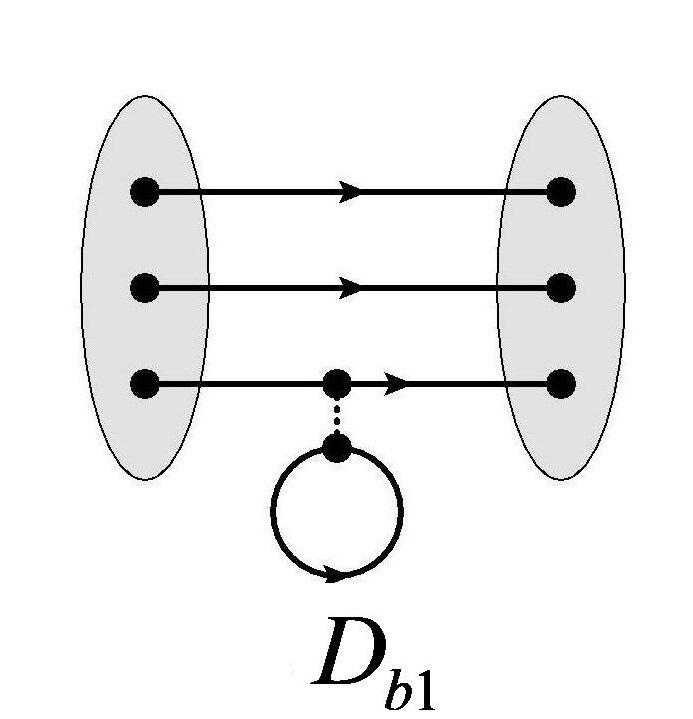}
		\includegraphics[scale=0.2]{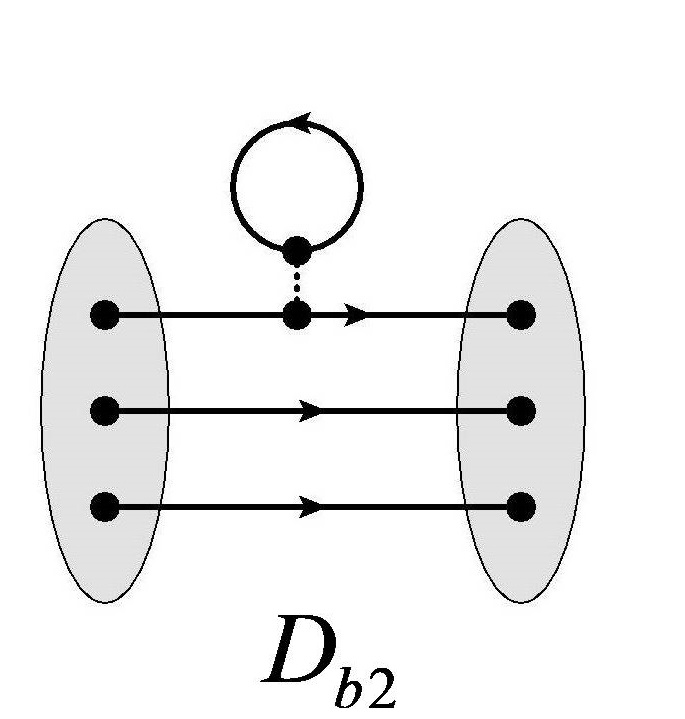}
		\includegraphics[scale=0.2]{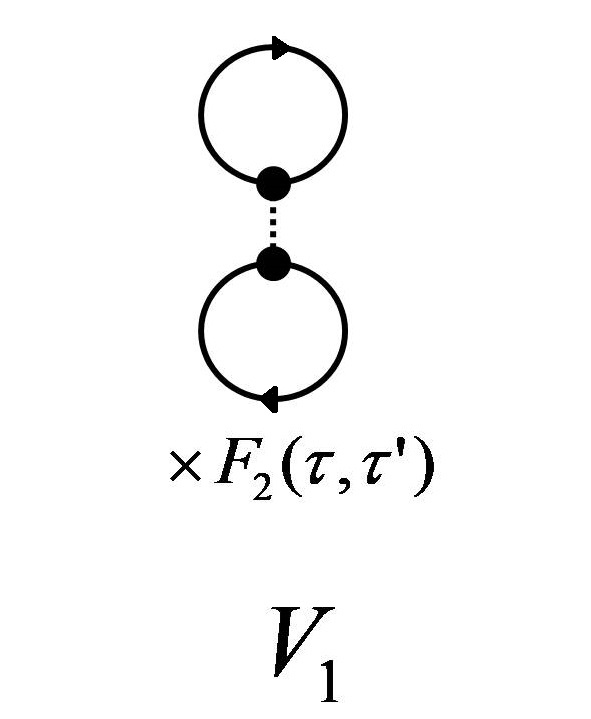}
		\includegraphics[scale=0.2]{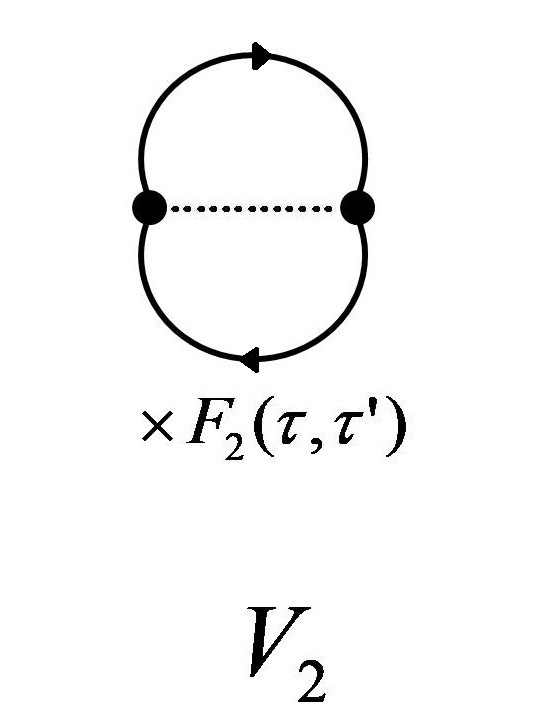}
		\par\end{centering}
	\caption{\label{fig:contract}List of contraction diagrams. {\color{black}In each diagram, the ellipses at the left and the right represent $\bar{\chi}(q_1,q_2,q_3;\vec{x}',\tau')$ and $\chi(q_1,q_2,q_3;\vec{x},\tau)$, respectively, whereas the solid lines represent the contractions between quark fields and the short dashed line denotes the four-quark vertex at $(\vec{x}_0,\tau_0)$.}}
\end{figure}

A contraction function $X(\tau,\tau_0,\tau')$ may now be defined as the contribution to the following general three-point function, 
\begin{equation}
\left\langle 0\right|\chi(q_1,q_2,q_3;\vec{x},\tau)\left[\bar{q}^a_i(\vec{x}_0,\tau_0)\Gamma^\mu q^{a}_j(\vec{x}_0,\tau_0)\bar{q}^{b}_{i'}(\vec{x}_0,\tau_0)\Gamma_\mu q^{b}_{j'}(\vec{x}_0,\tau_0)\right] \bar{\chi}(q_1,q_2,q_3;\vec{x}',\tau')\left|0\right\rangle ,
\end{equation}
where $\{i,i',j,j'\}$ are flavor indices, from a definite contraction diagram $X$. With the exchange symmetry of $\chi$ as shown in Eq.~\eqref{eq:flavorsymmetry}, it is straightforward to demonstrate that there are altogether eight types of independent contraction diagrams as depicted in Fig.~\ref{fig:contract}. In each diagram, the ellipse at the left represents $\bar{\chi}(q_1,q_2,q_3;\vec{x}',\tau')$, that at the right represents $\chi(q_1,q_2,q_3;\vec{x},\tau)$, whereas a solid line represents the contraction between a pair of quark fields and the short dashed line denotes the four-quark vertex. We shall name the first two diagrams as ``connected", the next four as ``quark loop" and the last two as ``vacuum" diagrams, respectively. One may read off the explicit form of each contraction function $X(\tau,\tau_0,\tau')$ from Fig.~\ref{fig:contract}. For instance, {\color{black}
\begin{eqnarray}
C_1(\tau,\tau_0,\tau')&=& 
\varepsilon^{abc}\varepsilon^{a'b'c'}\left\langle 0\right|\gamma_5\gamma^\mu
\bcontraction{}{q}{{}_1^aq_2^{b\mathrm{T}}C\gamma_\mu q_3^c \big[\bar{q}_2^d\Gamma^\alpha q_2^d\bar{q}_3^e\Gamma_\alpha q_3^e\big] \bar{q}_3^{c'}\gamma_\beta C\bar{q}_2^{b'\mathrm{T}}}{\bar{q}}
\acontraction[2ex]{q_1^a}{q}{{}_2^{b\mathrm{T}}C\gamma_\mu q_3^c\big[}{\bar{q}}
\acontraction[2ex]{q_1^aq_2^{b\mathrm{T}}C\gamma_\mu q_3^c\big[\bar{q}_2^d\Gamma^\alpha}{q}{{}_2^d\bar{q}_3^e\Gamma_\alpha q_3^e\big]\bar{q}_3^{c'}\gamma_\beta C}{\bar{q}}
\bcontraction[2ex]{q_1^aq_2^{b\mathrm{T}}C\gamma_\mu}{q}{{}_3^c\big[\bar{q}_2^d\Gamma^\alpha q_2^d}{\bar{q}}
\bcontraction[2ex]{q_1^aq_2^{b\mathrm{T}}C\gamma_\mu q_3^c \big[\bar{q}_2^d\Gamma^\alpha q_2^d\bar{q}_3^e\Gamma_\alpha}{q}{{}_3^e\big]}{\bar{q}}
q_1^aq_2^{b\mathrm{T}}C\gamma_\mu q_3^c\big[\bar{q}_2^d\Gamma^\alpha q_2^d\bar{q}_3^e\Gamma_\alpha q_3^e\big]\bar{q}_3^{c'}\gamma_\beta C\bar{q}_2^{b'\mathrm{T}}\bar{q}_1^{a'}\gamma_5\gamma^\beta\left|0\right\rangle\nonumber\\
D_{a1}(\tau,\tau_0,\tau')&=& 
\varepsilon^{abc}\varepsilon^{a'b'c'}\left\langle 0\right|\gamma_5\gamma^\mu
\bcontraction{}{q}{{}_1^aq_2^{b\mathrm{T}}C\gamma_\mu q_3^c \big[\bar{q}_3^d\Gamma^\alpha q_3^d\bar{q}_3^e\Gamma_\alpha q_3^e\big] \bar{q}_3^{c'}\gamma_\beta C\bar{q}_2^{b'\mathrm{T}}}{\bar{q}}
\acontraction[2ex]{q_1^a}{q}{{}_2^{b\mathrm{T}}C\gamma_\mu q_3^c\big[\bar{q}_3^d\Gamma^\alpha q_3^d\bar{q}_3^e\Gamma_\alpha q_3^e\big]\bar{q}_3^{c'}\gamma_\beta C}{\bar{q}}
\bcontraction[2ex]{q_1^aq_2^{b\mathrm{T}}C\gamma_\mu}{q}{{}_3^c\big[}{\bar{q}}
\bcontraction[2ex]{q_1^aq_2^{b\mathrm{T}}C\gamma_\mu q_3^c \big[\bar{q}_3^d\Gamma^\alpha} {q}{{}_3^d}{\bar{q}}
\bcontraction[2ex]{q_1^aq_2^{b\mathrm{T}}C\gamma_\mu q_3^c\big[\bar{q}_3^d\Gamma^\alpha q_3^d\bar{q}_3^e\Gamma_\alpha} {q}{{}_3^e\big]}{\bar{q}}
q_1^aq_2^{b\mathrm{T}}C\gamma_\mu q_3^c\big[\bar{q}_3^d\Gamma^\alpha q_3^d\bar{q}_3^e\Gamma_\alpha q_3^e\big]\bar{q}_3^{c'}\gamma_\beta C\bar{q}_2^{b'\mathrm{T}}\bar{q}_1^{a'}\gamma_5\gamma^\beta\left|0\right\rangle\nonumber\\
D_{b1}(\tau,\tau_0,\tau')&=& 
\varepsilon^{abc}\varepsilon^{a'b'c'}\left\langle 0\right|\gamma_5\gamma^\mu
\bcontraction{}{q}{{}_1^aq_2^{b\mathrm{T}}C\gamma_\mu q_3^c\big[\bar{q}_3^d\Gamma^\alpha q_3^d\bar{q}_4^e\Gamma_\alpha q_4^e\big]\bar{q}_3^{c'}\gamma_\beta C\bar{q}_2^{b'\mathrm{T}}}{\bar{q}}
\acontraction{q_1^a}{q}{{}_2^{b\mathrm{T}}C\gamma_\mu q_3^c\big[\bar{q}_3^d\Gamma^\alpha q_3^d\bar{q}_4^e\Gamma_\alpha q_4^e\big]\bar{q}_3^{c'}\gamma_\beta C}{\bar{q}}
\bcontraction[2ex]{q_1^aq_2^{b\mathrm{T}}C\gamma_\mu} {q}{{}_3^c\big[}{\bar{q}}
\bcontraction[2ex]{q_1^aq_2^{b\mathrm{T}}C\gamma_\mu q_3^c\big[\bar{q}_3^d\Gamma^\alpha} {q}{{}_3^d\bar{q}_4^e\Gamma_\alpha q_4^e\big]}{\bar{q}}
\acontraction[2ex]{q_1^aq_2^{b\mathrm{T}}C\gamma_\mu q_3^c\big[\bar{q}_3^d\Gamma^\alpha q_3^d}{\bar{q}}{{}_4^e\Gamma_\alpha}{q}
q_1^aq_2^{b\mathrm{T}}C\gamma_\mu q_3^c\big[\bar{q}_3^d\Gamma^\alpha q_3^d\bar{q}_4^e\Gamma_\alpha q_4^e\big]\bar{q}_3^{c'}\gamma_\beta C\bar{q}_2^{b'\mathrm{T}}\bar{q}_1^{a'}\gamma_5\gamma^\beta\left|0\right\rangle,
\end{eqnarray}}
where we have suppressed the spacetime coordinates: $(\vec{x},\tau)$ for the first three quark fields, $(\vec{x}_0,\tau_0)$ for the next four inside the square brackets and $(\vec{x}',\tau')$ for the last three. Notice also that in such a definition the contraction function $X(\tau,\tau_0,\tau')$ includes all possible minus signs due to the switching of positions between quark fields. For the vacuum contractions $V_1(\tau,\tau_0,\tau')$ and $V_2(\tau,\tau_0,\tau')$, we choose to normalize them according to the two-point function $F_2(\tau,\tau')$ which is independent of the choice of baryon in the flavor degenerate limit. With these, we may now define the quantity $M_X$ as
\begin{equation}
M_X=2m_NL^3\lim_{\substack{\tau-\tau_0\rightarrow +\infty\\\tau_0-\tau'\rightarrow+\infty}}\frac{ X(\tau,\tau_0,\tau')}{F_2(\tau,\tau')}.\label{eq:defMX}
\end{equation}
We will show that the limits exist in a short while. The matrix element of any four-quark operator with respect to any external spin-1/2 static baryon states can now be expressed as a linear combination of $\{M_X\}$. For instance, 
\begin{equation}
\left\langle p\right|\hat{O}\left|p\right\rangle=M_{C_1}+2\left(M_{D_{a1}}+M_{D_{b1}}\right)-\left(M_{D_{a2}}+M_{D_{b2}}\right).\label{eq:Omatrix}
\end{equation}
This is how one could systematically express a four-quark matrix element in terms of contributions from different contraction diagrams on lattice. 

In practice, the lattice computations of different types of $X(\tau,\tau_0,\tau')$ contractions will involve very different techniques. While the $C_i(\tau,\tau_0,\tau')$ can usually be computed quite economically with high precision, the quark loop contractions $D_{ai}(\tau,\tau_0,\tau')$ and $D_{bi}(\tau,\tau_0,\tau')$ contain a quark propagator that starts and ends at the same spacetime point and is extremely noisy. For such propagators, one needs to average its value over all lattice points in order to improve its signal, but then it requires the use of all-to-all propagators that are computationally expensive~\cite{Bitar:1988bb,Bernardson:1994at,Kuramashi:1993ka,Dong:1993pk,Eicker:1996gk,deDivitiis:1996qx,Michael:1998sg,McNeile:2000xx,Neff:2001zr,DeGrand:2002gm,Duncan:2001ta,Bali:2005fu,Foley:2005ac,Boucaud:2008xu,Bali:2009hu}. As a consequence, with a given computational power, the precision level for the lattice outcomes of $C_i$ and $\{D_{ai},D_{bi}\}$ are very different. Of course, the physical results require summing all of them in the way given in Eq.~\eqref{eq:Omatrix} for the proton. However, we may employ the ability of calculating contractions separately on lattice as a handle to improve the precision of the final results. For that we can carry out the chiral extrapolations for different types of contractions separately, and this requires the analytic expression of each $M_X$ as a function of the pion mass.

In the physical world it is usually not possible to separate the connected contractions from the quark loops in a given matrix element. Such separation is however possible in a QCD with an extended flavor sector. Let us consider a strong interaction theory with four fermionic quarks $\{u,d,j,k\}$ and two bosonic ``ghost" quarks $\{\tilde{j},\tilde{k}\}$ with degenerate masses, which can be written collectively as $q'=(u\:\:d\:\:j\:\:k\:|\:\tilde{j}\:\:\tilde{k})^T$. All internal dynamics of such a theory will be identical to the ordinary two-flavor QCD because all the loop effects brought up by the two extra fermionic quarks $\{j,k\}$ (known as ``valence quarks") are exactly canceled by their corresponding bosonic partners $\{\tilde{j},\tilde{k}\}$, keeping the sea DOFs unchanged. The net effect of this extension is that one introduces quark flavors that can only appear in external states but not in loops, and the strong interaction theory of such system is known as the partially-quenched QCD (PQQCD)~\cite{Bernard:1993sv,Sharpe:1999kj}. Within this framework, any contraction diagram of interest can be constructed by appropriately choosing the quark contents in either the external states or the operators~\cite{Sharpe:2006pu}; such an idea has been previously applied in studies of the hadronic vacuum polarization~\cite{DellaMorte:2010aq}, the pion scalar form factor~\cite{Juttner:2011ur} and the $\pi\pi$ scattering amplitudes~\cite{Acharya:2017zje}. In our case, one could easily demonstrate that each quantity $M_X$ can be written as a linear combination of four-quark matrix elements in PQQCD; an explicit example is given in Appendix~\ref{sec:MESU42}. With that we also show that each contraction function $X(\tau,\tau_0,\tau')$ has the correct asymptotic exponential behavior of $\exp\{-m_N(\tau-\tau')\}$ that guarantees the existence of the limits in Eq.~\eqref{eq:defMX}.

\section{PQChPT analysis\label{sec:PQChPTanalysis}}

In the previous section we have successfully separated each contraction diagram into well-defined matrix elements in PQQCD, and here we shall proceed to study the low-energy behavior of each individual contraction that contributes to the four-quark matrix element $\left\langle p\right|\hat{O}\left|p\right\rangle$ of our interest. This involves the application of the low-energy EFT of PQQCD as follows.

In the massless limit, PQQCD with four fermionic and two ghost quarks has a ``graded" SU(4$|$2) chiral symmetry, namely the Lagrangian is invariant under the transformation
\begin{equation}q'_{R}\rightarrow Rq'_{R},\:\:\:q'_{L}\rightarrow Lq'_{L} ,
\end{equation}
where $R\in\mathrm{SU(4}|2)_{R}$,  $L\in\mathrm{SU(4}|2)_{L}$ are elements of a special unitary (4$|$2) graded symmetry group. At low energy, this graded chiral symmetry is spontaneously broken as $\mathrm{SU(4}|2)_L\times\mathrm{SU(4}|2)_R\rightarrow \mathrm{SU}(4|2)_V$ which generates 35 pNG particles in complete analogy to the ordinary chiral symmetry breaking of two-flavor QCD: $\mathrm{SU}(2)_L\times \mathrm{SU}(2)_R\rightarrow \mathrm{SU}(2)_V$. One may proceed to write down the PQChPT~\cite{Sharpe:2000bc,Sharpe:2001fh,Sharpe:2006pu,Golterman:2009kw} which is the low-energy EFT pf PQQCD that incorporates the interactions between the pNG particles and other matter fields. 

The essential information of heavy baryon PQChPT can be found in Appendix~\ref{sec:HBPQxPT} and the references listed above, so here we shall concentrate on the implementation of the weak interaction in the chiral Lagrangian. Let us consider a generic four-quark operator in $\mathrm{SU}(4|2)$,
\begin{equation}
\hat{O}_{4q}=\bar{q}^{\prime a}\gamma^\mu \tau_\mathcal{A} q^{\prime a}\bar{q}^{ \prime b}\gamma_\mu \tau_\mathcal{B} q^{\prime b}, \label{eq:O4qdef}
\end{equation}
where $\tau_\mathcal{A},\tau_\mathcal{B}$ are $\mathrm{SU}(4|2)$ generators; we shall assume $\hat{O}_{4q}$ to be Hermitian for simplicity. 
One may decompose the quark fields in $\hat{O}_{4q}$ into left- and right-handed components and obtain
\begin{equation}
\hat{O}_{4q}=\left(\bar{q}^{\prime a}_R\gamma^\mu \tau_\mathcal{A} q^{\prime a}_R+\bar{q}^{ \prime a}_L\gamma^\mu \tau_\mathcal{A} q^{\prime a}_L\right)\left(\bar{q}^{\prime b}_R\gamma_\mu \tau_\mathcal{B} q^{\prime b}_R+\bar{q}^{\prime b}_L\gamma_\mu \tau_\mathcal{B} q^{\prime b}_L\right).
\end{equation}
It is now obvious that the implementation of $\hat{O}_{4q}$ in the chiral Lagrangian will involve spurions of the form $\tilde{X}^\mathcal{A}\otimes \tilde{X}^\mathcal{B}+(\mathcal{A}\leftrightarrow \mathcal{B})$, where $\tilde{X}^{\mathcal{A},\mathcal{B}}=X^{\mathcal{A},\mathcal{B}}_R+X^{\mathcal{A},\mathcal{B}}_L=u^\dagger\tau_{\mathcal{A},\mathcal{B}}u+u\tau_{\mathcal{A},\mathcal{B}}u^\dagger$, which is a straightforward generalization of the spurion introduced in Section~\ref{sec:hpimatch}. With this, there are eight independent operators in the baryon sector one could write down at LO,
\begin{eqnarray}
\hat{O}_{C_1}&=&\frac{1}{2}\bar{B}_{k'j'i}\tilde{X}^\mathcal{A}_{k'k}\tilde{X}^\mathcal{B}_{j'j}B_{ijk}(-1)^{(\eta_i+\eta_{j'})(\eta_k+\eta_{k'})+\eta_i(\eta_j+\eta_{j'})}+(\mathcal{A}\leftrightarrow\mathcal{B}) , \nonumber\\
\hat{O}_{C_2}&=&\frac{1}{2}\bar{B}_{kj'i'}\tilde{X}^\mathcal{A}_{j'j}\tilde{X}^\mathcal{B}_{i'i}B_{ijk}(-1)^{\eta_{i'}(\eta_j+\eta_{j'})}+(\mathcal{A}\leftrightarrow\mathcal{B}) , \nonumber\\
\hat{O}_{D_{a1}}&=&\frac{1}{2}\bar{B}_{k'ji}\tilde{X}^\mathcal{A}_{k'l}\tilde{X}^\mathcal{B}_{lk}B_{ijk}(-1)^{(\eta_i+\eta_j)(\eta_k+\eta_{k
		'})}+(\mathcal{A}\leftrightarrow\mathcal{B}) , \nonumber\\
\hat{O}_{D_{a2}}&=&\frac{1}{2}\bar{B}_{kji'}\tilde{X}^\mathcal{A}_{i'l}\tilde{X}^\mathcal{B}_{li}B_{ijk}+(\mathcal{A}\leftrightarrow\mathcal{B}) , \nonumber\\
\hat{O}_{D_{b1}}&=&\frac{1}{2}\bar{B}_{k'ji}\tilde{X}^\mathcal{A}_{k'k}B_{ijk}\mathrm{Str}[\tilde{X}^\mathcal{B}](-1)^{(\eta_k+\eta_{k'})(\eta_i+\eta_j)}+(\mathcal{A}\leftrightarrow\mathcal{B}) , \nonumber\\
\hat{O}_{D_{b2}}&=&\frac{1}{2}\bar{B}_{kji'}\tilde{X}^\mathcal{A}_{i'i}B_{ijk}\mathrm{Str}[\tilde{X}^\mathcal{B}]+(\mathcal{A}\leftrightarrow\mathcal{B}) , \nonumber\\
\hat{O}_{V_1}&=&\mathrm{Str}[\tilde{X}^\mathcal{A}]\mathrm{Str}[\tilde{X}^\mathcal{B}]\bar{B}_{kji}B_{ijk} , \nonumber\\
\hat{O}_{V_2}&=&\mathrm{Str}[\tilde{X}^\mathcal{A}\tilde{X}^\mathcal{B}]\bar{B}_{kji}B_{ijk} ,
\end{eqnarray}
where
\begin{equation}
\mathrm{Str}[A]=\sum_{i=1}^4A_{ii}-\sum_{i=5}^6A_{ii}.\label{eq:Str}
\end{equation}
The corresponding Lagrangian can be written as
\begin{eqnarray}
\mathcal{L}_{4q,\mathrm{LO}}&=& 
\alpha_{C_1}\hat{O}_{C_1}+\alpha_{C_2}\hat{O}_{C_2}+\alpha_{D_{a1}}\hat{O}_{D_{a1}}+\alpha_{D_{a2}}\hat{O}_{D_{a2}}+\alpha_{D_{b1}}\hat{O}_{D_{b1}}+\alpha_{D_{b2}}\hat{O}_{D_{b2}} \nonumber\\
&& +\alpha_{V_1}\hat{O}_{V_1}+\alpha_{V_2}\hat{O}_{V_2} . 
\label{eq:L4qLO}
\end{eqnarray}
Here we choose to label a given operator according to the name of the contraction diagram in Fig.~\ref{fig:contract} that contracts the quark indices in the same way as the operator. We observe from Eqs.~\eqref{eq:flavorsymmetry} and \eqref{eq:exchange} that the interpolating operator $\chi(q_1,q_2,q_3)$ and the PQChPT spin-1/2 baryon field $B_{ijk}$ share the same exchange symmetry relations when all quarks are fermionic (which is part of the reasons we choose this particular definition of $\chi$), so the number of independent operators at LO is exactly the same as the number of independent contraction diagrams. It is also important to point out that the LECs $\{\alpha_i\}$ are universal constants that do not depend on the actual choice of $\hat{O}_{4q}$, hence we are free to choose any form of $\tau_A$ and $\tau_B$ to arrive at our desired contraction diagrams. For instance, we show in Appendix~\ref{sec:MESU42} that the tree-level results for all eight independent contractions can be obtained by computing the proton matrix element with different choices of $\hat{O}_{4q}$.

\subsection{Tree-level and one-loop results}

With the LO Lagrangian in Eq.~\eqref{eq:L4qLO} one can now proceed to compute the tree and one-loop contributions to each $M_X$. Recall that our objective is to determine the matrix element $\left\langle p\right|\hat{O}\left|p\right\rangle$ which depends only on three combinations of contractions: $M_{C_1}$, $M_{D_{a1}}+M_{D_{b1}}$ and $M_{D_{a2}}+M_{D_{b2}}$ as indicated in Eq.~\eqref{eq:Omatrix}. There is an easier way to obtain their expressions than the general procedure depicted in Appendix~\ref{sec:MESU42}, namely, we fix the four-quark operator as $\hat{O}_{4q}=\hat{O}$, but choose different external baryon states. For instance, we define the $\mathrm{SU}(4|2)$ baryons $\tilde{\Sigma}^0$ and $\tilde{\Lambda}$ as the direct analogy to the physical $\Sigma^0$ and $\Lambda$ baryons, with the replacements $d\rightarrow j$ and $s\rightarrow k$. One may then show that
\begin{eqnarray}
\langle \tilde{\Sigma}^0|\hat{O}|\tilde{\Sigma}^0\rangle&=&M_{D_{a1}}+M_{D_{b1}}, \nonumber\\
\langle \tilde{\Lambda}|\hat{O}|\tilde{\Lambda}\rangle&=&\frac{1}{3}\left(M_{D_{a1}}+M_{D_{b1}}\right)+\frac{2}{3}\left(M_{D_{a2}}+M_{D_{b2}}\right). \label{eq:MESigmaLambda}
\end{eqnarray}
Therefore, a combination of Eqs.~\eqref{eq:Omatrix} and \eqref{eq:MESigmaLambda} allows a re-expression of $M_{C_1}$, $M_{D_{a1}}+M_{D_{a2}}$ and $M_{D_{b1}}+M_{D_{b2}}$ in terms of the diagonal matrix element of $\hat{O}$ with respect to the baryon states $p$, $\tilde{\Sigma}^0$ and $\tilde{\Lambda}$. One may then compute the latter to one loop using HB PQChPT to obtain the former. Below we summarize the main results. 
First, at tree-level, we obtain
\begin{eqnarray}
\left\langle p\right|\hat{O}\left|p\right\rangle_\mathrm{tree}&=&\frac{4}{3}m_N\left(4\alpha_{C_1}+\alpha_{C_2}+4\alpha_{D_1}-2\alpha_{D_2}\right), \nonumber\\
\langle \tilde{\Sigma}^0|\hat{O}|\tilde{\Sigma}^0\rangle_\mathrm{tree}&=&\frac{2}{3}m_N\left(5\alpha_{D_1}+2\alpha_{D_2}\right), \nonumber\\
\langle \tilde{\Lambda}|\hat{O}|\tilde{\Lambda}\rangle_\mathrm{tree}&=&2m_N\left(\alpha_{D_1}+2\alpha_{D_2}\right),
\end{eqnarray}
where we have defined
\begin{equation} \alpha_{D_1}=\alpha_{D_{a1}}+\alpha_{D_{b1}},\qquad \alpha_{D_2}=\alpha_{D_{a2}}+\alpha_{D_{b2}} 
\end{equation}
for compactness. 

Next we consider the one-loop contributions which can be further sub-divided into 1PI contributions and the wavefunction renormalization. There are two types of 1PI diagrams with topologies identical to those of Fig.~\ref{fig:hpimNloop} (e) and  (f), and their contributions are labeled as 1PI(e) and 1PI(f), respectively. The results are
\begin{eqnarray}
\delta \left\langle p\right|\hat{O}\left|p\right\rangle_\mathrm{1PI(e)}&=&\frac{4m_N}{3F_\pi^2}I_e\left(-4\alpha_{C_1}-\alpha_{C_2}-4\alpha_{D_1}+2\alpha_{D_2}\right), \nonumber\\
\delta \langle \tilde{\Sigma}^0|\hat{O}|\tilde{\Sigma}^0\rangle_\mathrm{1PI(e)}&=&\frac{m_N}{3F_\pi^2}I_e\left(10\alpha_{C_1}+7\alpha_{C_2}-10\alpha_{D_1}-4\alpha_{D_2}\right), \nonumber\\
\delta \langle \tilde{\Lambda}|\hat{O}|\tilde{\Lambda}\rangle_\mathrm{1PI(e)}&=&\frac{m_N}{F_\pi^2}I_e\left(2\alpha_{C_1}+3\alpha_{C_2}-2\alpha_{D_1}-4\alpha_{D_2}\right),
\end{eqnarray}
and
\begin{eqnarray}
\delta \left\langle p\right|\hat{O}\left|p\right\rangle_\mathrm{1PI(f)}&=&\frac{4m_N}{27F_\pi^2}I_a\left(\beta-2\rho\right)^2\left(4\alpha_{C_1}+\alpha_{C_2}+4\alpha_{D_1}-2\alpha_{D_2}\right), \nonumber\\
\delta \langle \tilde{\Sigma}^0|\hat{O}|\tilde{\Sigma}^0\rangle_\mathrm{1PI(f)}&=&\frac{2m_N}{27F_\pi^2}I_a\left[-(4\alpha_{C_1}+\alpha_{C_2})(10\beta^2+14\beta\rho+13\rho^2)+\alpha_{D_1}(5\beta^2+34\beta\rho-7\rho^2)\right.\nonumber\\
&&\left.+\alpha_{D_2}(38\beta^2+64\beta\rho-37\rho^2)\right], \nonumber\\
\delta \langle \tilde{\Lambda}|\hat{O}|\tilde{\Lambda}\rangle_\mathrm{1PI(f)}&=&\frac{2m_N}{9F_\pi^2}I_a\left[-(4\alpha_{C_1}+\alpha_{C_2})(2\beta^2-2\beta\rho+5\rho^2)+\alpha_{D_1}(\beta^2+26\beta\rho-11\rho^2)\right.\nonumber\\
&&\left.+\alpha_{D_2}(22\beta^2+32\beta\rho-17\rho^2)\right].
\end{eqnarray}
Finally, the wavefunction renormalization $Z_N$ is identical to that in the ordinary two-flavor ChPT result in Eq.~\eqref{eq:ZN} and it applies to all fermionic baryons due to the exact $\mathrm{SU}(4|2)_V$ symmetry.

The one-loop calculations above contain UV divergences which need to be canceled by counterterms of order $\mathcal{O}(m_\pi^2/\Lambda_\chi^2)$. Here we shall not bother to write down the full counterterm Lagrangian because there are too many available terms, and eventually what we only need to know is that $M_{C_1}$, $M_{D_{a1}}+M_{D_{b1}}$ and $M_{D_{a2}}+M_{D_{b2}}$ will acquire independent combinations of counterterms. Therefore, by collecting tree-level, one-loop and counterterm contributions, we obtain
\begin{eqnarray}
M_{C_1}&=&\frac{4}{3}m_N\left(4\alpha_{C_1}+\alpha_{C_2}\right)\left[1+\frac{1}{3F_\pi^2}I_a\left(4\beta^2+2\beta\rho+7\rho^2\right)-\frac{2}{F_\pi^2}I_e\right]+\frac{m_\pi^2}{\Lambda_\chi^2}\delta\alpha_{C_1}m_N , \nonumber\\
M_{D_{a1}}+M_{D_{b1}}&=&\frac{2}{3}m_N\left(5\alpha_{D_1}+2\alpha_{D_2}\right)+\frac{2m_N}{27F_\pi^2}I_a\left[-(4\alpha_{C_1}+\alpha_{C_2})(10\beta^2+14\beta\rho+13\rho^2)\right.\nonumber\\
&&\left.+\alpha_{D_1}(20\beta^2-26\beta\rho+53\rho^2)+\alpha_{D_2}(44\beta^2+40\beta\rho-13\rho^2)\right]\nonumber\\
&&+\frac{m_N}{3F_\pi^2}I_e\left(10\alpha_{C_1}+7\alpha_{C_2}-10\alpha_{D_1}-4\alpha_{D_2}\right)+\frac{m_\pi^2}{\Lambda_\chi^2}\delta\alpha_{D_1}m_N , \nonumber\\
M_{D_{a2}}+M_{D_{b2}}&=&\frac{4}{3}m_N\left(\alpha_{D_1}+4\alpha_{D_2}\right)+\frac{4m_N}{27F_\pi^2}I_a\left[-2(4\alpha_{C_1}+\alpha_{C_2})(\beta-2\alpha)^2\right.\nonumber\\
&&\left.+\alpha_{D_1}(4\beta^2+38\beta\rho-11\rho^2)+\alpha_{D_2}(52\beta^2+8\beta\rho+19\rho^2)\right]\nonumber\\
&&+\frac{2m_N}{3F_\pi^2}I_e\left(2\alpha_{C_1}+5\alpha_{C_2}-2\alpha_{D_1}-8\alpha_{D_2}\right)+\frac{m_\pi^2}{\Lambda_\chi^2}\delta\alpha_{D_2}m_N.\label{eq:Mfull}
\end{eqnarray}
Here $m_N$ is the physical nucleon mass that has the full quark-mass dependence, and $\{\delta\alpha_{C_1},\delta\alpha_{D_1},\delta\alpha_{D_2}\}$ represent the total counterterm contribution to each respective quantity.

\begin{figure}[t]
	\begin{centering}
		\includegraphics[scale=0.13]{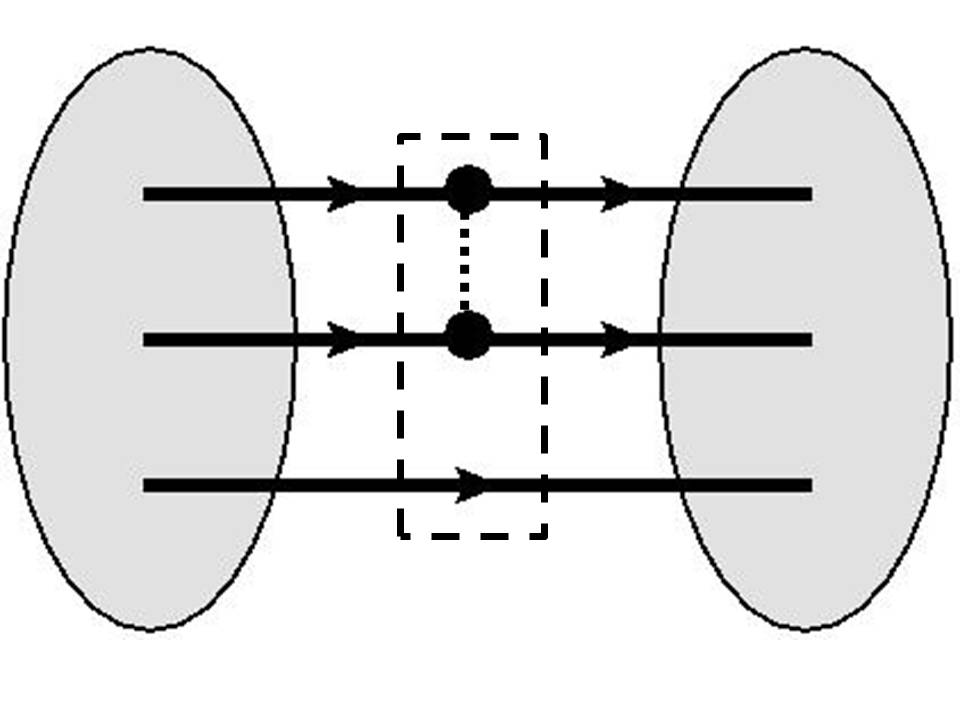}\:\:
		\includegraphics[scale=0.13]{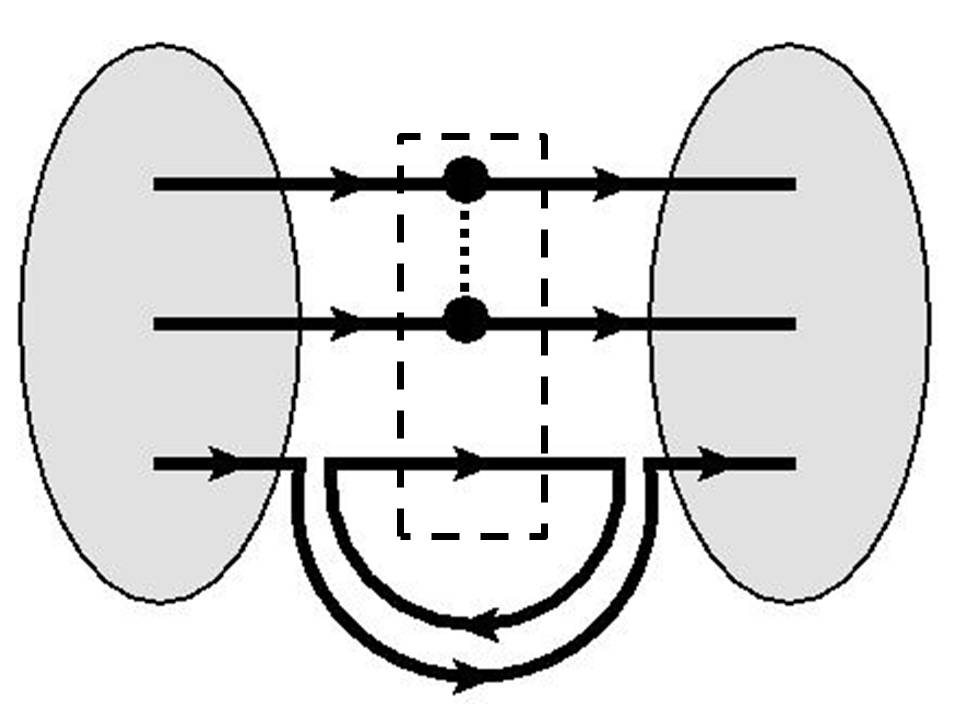}\:\:
		\includegraphics[scale=0.13]{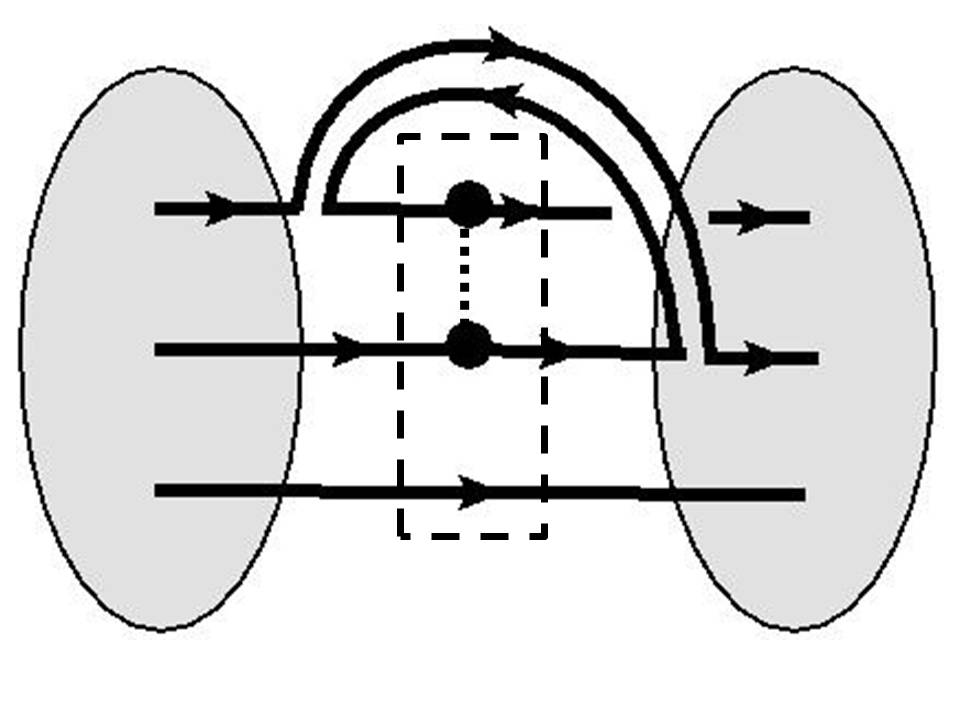}\:\:
		\includegraphics[scale=0.13]{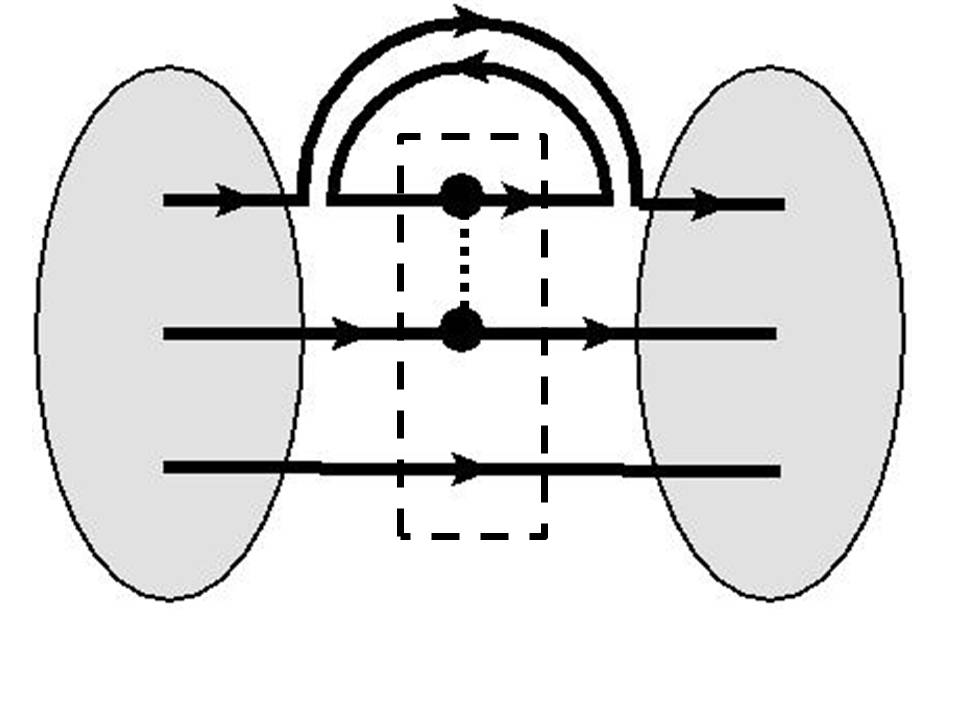}
		\par\end{centering}
	\caption{\label{fig:induce}Illustration of how a short-distance connected four quark interaction, represented by the interaction vertex in the dashed box, could induce (1) connected diagram (the second), (2) quark-loop diagram type $D_a$ (the third) and (3) quark-loop diagram type $D_b$ (the fourth) under long-range QCD corrections.}
\end{figure}

Let us try to understand the results above. The first important observation is that the two LECs $\{\alpha_{C_1},\alpha_{C_2}\}$, which contribute only to connected diagrams at tree level, enter the quark loop diagrams in the form of chiral logarithms; this feature can be understood diagrammatically as depicted in Fig.~\ref{fig:induce}. On the other hand, there is no way that the LECs $\{\alpha_{D_i}\}$ can induce connected diagrams through loop corrections, and therefore we observe that both the tree-level and chiral logarithms of $M_{C_1}$ depend only on $\{\alpha_{C_i}\}$. Since connected diagrams can be readily computed on lattice, they may be computed with several values of $m_\pi$ which then, using our derived formula, allow for a determination of the LECs $\{\alpha_{C_i}\}$. By doing so, we do not just acquire the full information of the connected diagrams, but also fix a part of the chiral logarithms in the quark loop diagrams. Even though the leading terms and the remaining chiral logarithms of the latter can only be fixed by direct lattice calculations of such diagrams, now they depend on a smaller amount of unknown LECs (i.e., just $\alpha_{D_1}$ and $\alpha_{D_2}$), making their chiral extrapolation much easier. 

We should end this section by mentioning another technical detail. From Eq.~\eqref{eq:Mfull} we find that the calculation of $M_{C_1}$ can only fix the combination $4\alpha_{C_1}+\alpha_{C_2}$ but we do need to know the two LECs separately. Therefore, we shall supply also the theoretical formula for the other contraction diagram $M_{C_2}$. The easiest way to calculate it is to realize that if we choose $\hat{O}'=\bar{q}^{\prime a}\gamma^\mu q^{\prime a}(\bar{u}^b\gamma_\mu u^b-\bar{d}^b\gamma_\mu d^b)$, then we have
\begin{equation}
\langle \tilde{\Sigma}^0|\hat{O}'|\tilde{\Sigma}^0\rangle=\frac{1}{2}M_{C_1}+\frac{1}{2}M_{C_2}+M_{D_{a1}}+M_{D_{b1}},
\end{equation}
and thus we only need to compute one more matrix element $\langle \tilde{\Sigma}^0|\hat{O}'|\tilde{\Sigma}^0\rangle$. We simply quote the final result,
\begin{eqnarray}
M_{C_2}&=&\frac{2}{3}m_N\left(2\alpha_{C_1}+5\alpha_{C_2}\right)+\frac{4m_N}{9F_\pi^2}I_a\left[\alpha_{C_1}(4\beta^2+2\beta\rho+7\rho^2)+\alpha_{C_2}(10\beta^2+5\beta\rho+4\rho^2)\right]\nonumber\\
&&-\frac{4m_N}{3F_\pi^2}I_e\left(2\alpha_{C_1}+5\alpha_{C_2}\right)+\frac{m_\pi^2}{\Lambda_\chi^2}\delta\alpha_{C_2}m_N.
\end{eqnarray}
Thus, by computing $M_{C_1}$ and $M_{C_2}$ on lattice one is able to fix $\alpha_{C_1}$ and $\alpha_{C_2}$ simultaneously.

\section{Spin-flavor symmetry}

In Section~\ref{sec:PQChPTanalysis}, we show that in order to determine the tree-level and chiral logarithmic terms in $\left\langle p\right|\hat{O}\left|p\right\rangle$ it is necessary to perform a global fit of the pion mass dependence using four LECs: $\{\alpha_{C_1},\alpha_{C_2},\alpha_{D_1},\alpha_{D_2}\}$.
It would be beneficial to obtain approximate relations among the LECs, especially between the more difficult $\alpha_{D_1}$ and $\alpha_{D_2}$ in order to provide a guidance to the starting point of the global fit. To facilitate such a discussion, we start by providing another matrix representation of the combination $M_{D_{a1}}+M_{D_{b1}}$ and $M_{D_{a2}}+M_{D_{b2}}$. If we choose the four-quark operator,
\begin{eqnarray}
\hat{O}_{uk}&=&\hat{O}_{uu}-\hat{O}_{kk}\nonumber\\
&=&\bar{u}^a\gamma^\mu u^a\bar{u}^b\gamma_\mu u^b-(u\rightarrow k),
\end{eqnarray} 
then we have
\begin{eqnarray}
\langle \tilde{\Sigma}^0|\hat{O}_{uk}|\tilde{\Sigma}^0\rangle &=&\left(M_{D_{a1}}+M_{D_{b1}}\right)-\left(M_{D_{a2}}+M_{D_{b2}}\right), \nonumber\\
\left\langle n\right|\hat{O}_{uk}\left|n\right\rangle &=&\left(M_{D_{a2}}+M_{D_{b2}}\right).\label{eq:OukME}
\end{eqnarray}

Approximate relations between the matrix elements above can then be obtained by considering the spin-flavor symmetry~\cite{PaisSU6} among baryons. In a QCD with $N$ fermionic quark flavors, such a symmetry means that the quarks $\{q_i\}$ of a definite flavor $q$ and spin $i$ form the fundamental representation of a $\mathrm{SU}(2N)$ group and all hadrons can be grouped into irreducible representations of that symmetry group. For instance, the baryon octet and decuplet collectively form an irreducible 56-plet of the spin-flavor $\mathrm{SU}(6)$. It is well-known that the spin-flavor symmetry is a direct consequence of the large-$N_c$ limit (with $N_c$ the number of colors), and for most of the practical purposes it is simply equivalent to the non-relativistic quark model~\cite{Dashen:1994qi}. Therefore, for the discussion here let us consider the spin-flavor wave function of a spin-up $\tilde{\Sigma}^0$ state in the quark-model (QM) representation,
\begin{equation}
|\tilde{\Sigma}^0\rangle_\mathrm{QM}=\frac{\varepsilon^{abc}}{\sqrt{36}}\{\hat{k}_\uparrow^{a\dagger}\hat{j}_\downarrow^{b\dagger}\hat{u}_\uparrow^{c\dagger}+\hat{k}_\uparrow^{a\dagger}\hat{j}_\uparrow^{b\dagger}\hat{u}_\downarrow^{c\dagger}-2\hat{k}_\downarrow^{a\dagger}\hat{j}_\uparrow^{b\dagger}\hat{u}_\uparrow^{c\dagger}\}|0\rangle ,
\end{equation} 
where the quark creation and annihilation operators satisfy the anti-commutation relation
\begin{equation}
\{\hat{q}_i^a,\hat{q}_j^{b\dagger}\}=\delta_{ab}\delta_{ij} ,
\end{equation}
and the baryon state is normalized as $_\mathrm{QM}\langle\tilde{\Sigma}^0|\tilde{\Sigma}^0\rangle_\mathrm{QM}=1$. With this, we can compute the matrix element of $\hat{O}_{uu}$ and $\hat{O}_{kk}$ with respect to the spin-up $\tilde{\Sigma}^0$ state,
\begin{eqnarray}
_\mathrm{QM}\langle\tilde{\Sigma}^0_\uparrow|\hat{O}_{uu}|\tilde{\Sigma}^0_\uparrow\rangle_\mathrm{QM}&=&\frac{5}{36}\varepsilon^{abc}\varepsilon^{abc'}\langle u_\uparrow^c|\hat{O}_{uu}|u_\uparrow^{c'}\rangle+\frac{1}{36}\varepsilon^{abc}\varepsilon^{abc'}\langle u_\downarrow^c|\hat{O}_{uu}|u_\downarrow^{c'}\rangle\nonumber\\
&=&\frac{1}{6}\varepsilon^{abc}\varepsilon^{abc'}\langle u_\uparrow^c|\hat{O}_{uu}|u_\uparrow^{c'}\rangle.\label{eq:OuuME}
\end{eqnarray}
The last equality is due to rotational symmetry. Similarly, one also obtains
\begin{equation}
_\mathrm{QM}\langle\tilde{\Sigma}^0_\uparrow|\hat{O}_{kk}|\tilde{\Sigma}^0_\uparrow\rangle_\mathrm{QM}=\frac{1}{6}\varepsilon^{abc}\varepsilon^{abc'}\langle k_\uparrow^c|\hat{O}_{kk}|k_\uparrow^{c'}\rangle.\label{eq:OkkME}
\end{equation}
Therefore, combining Eqs.~\eqref{eq:OuuME}, \eqref{eq:OkkME} and the $u\leftrightarrow k$ flavor symmetry, we arrive at $_\mathrm{QM}\langle\tilde{\Sigma}^0_\uparrow|\hat{O}_{uk}|\tilde{\Sigma}^0_\uparrow\rangle_\mathrm{QM}=0$ which implies the following approximate relation according to Eq. \eqref{eq:OukME},
\begin{equation} 
M_{D_{a1}}+M_{D_{b1}}\approx M_{D_{a2}}+M_{D_{b2}},
\end{equation}
as a consequence of the spin-flavor symmetry. At LO in HB PQChPT, this in turns implies an approximate relation to the LECs,
\begin{equation}
\alpha_{D_1}\approx 2\alpha_{D_2}.\label{eq:alphaDSF}
\end{equation}
The equation above may now serve as a starting point for the global fit of the four LECs $\{\alpha_{C_1},\alpha_{C_2},\alpha_{D_1},\alpha_{D_2}\}$.

We end this section by commenting on the spin-flavor symmetry at one loop. We observe that the one-loop corrections of type 1PI(e) preserve the spin-flavor symmetry while those of type 1PI(f) do not. The reason is that under the spin-flavor symmetry the spin-1/2 and 3/2 baryons belong to the same multiplet and thus have to be taken simultaneously as dynamical DOFs. Our treatment of Fig.~\ref{fig:hpimNloop}(f), however, includes only spin-1/2 baryons whereas the effects of the rest get buried in the counterterms. This results in the explicit breaking of the symmetry in this particular diagram. 

\section{Operators with strange quarks}

In this section we shall briefly comment on the proton matrix elements of the four-quark operators with strange quark fields, i.e., $\{\tilde{\theta}_i^{(s)\prime}\}$ in Eq.~\eqref{eq:4qeven}. They give rise to contraction diagrams of type $D_{b}$, except that now the quark in the loop has a heavier mass. It is unavoidable that these quark loop diagrams must be calculated directly on lattice if we are to study their contributions to $h_\pi^1$. Nevertheless, the EFT analysis is still beneficial as it provides an extrapolation formula of the matrix element with respect to the light quark mass, which can be read off directly from Eq.~\eqref{eq:deltahpimN}: 
\begin{equation}
\left\langle p\right|\theta_i^{(s)\prime}\left|p\right\rangle =2m_N\alpha_i^{(s)}\left(1+\frac{4g_0^2}{F_\pi^2}I_a-\frac{1}{F_\pi^2}I_e\right)+\frac{m_\pi^2}{\Lambda_\chi^2}\delta\alpha_i^{(s)}m_N.
\end{equation}
Notice that both the LO term and the chiral logarithms depend only on one single LEC $\alpha_i^{(s)}$. One could therefore compute this matrix element on lattice with unphysical pion mass, which is presumably easier, and then extrapolate the result to the physical region using the formula above.

\section{Conclusions}

HPV has been studied for many years.
Among others, the $\Delta I=1$ HPV holds a special role as a unique probe of hadronic neutral weak current as well as one of the main contributors of of long-range nuclear PV. Moreover, with the release of the NPDGamma result, the $\Delta I=1$ P-odd pion-nucleon coupling $h_\pi^1$ is now the only DDH coupling with a definite isospin that has been numerically measured through a single experiment. Therefore, the first-principle calculations of $h_\pi^1$ are highly desirable as they are directly comparable to experimental results.

Despite the above, currently we observe a lack of progress in the lattice study of the $\Delta I=1$ HPV comparing to its $\Delta I=2$ counterpart. The latter involves a direct computation of nucleon-nucleon scattering amplitudes with the insertion of $\Delta I=2$ four-quark operators. It does not require computations of noisy disconnected diagrams but the total amount of contractions is tremendous even in the exact isospin limit. On the other hand, although disconnected diagrams are unavoidable in the $\Delta I=1$ channel, the total amount of contractions is much less. Furthermore, given its more straightforward relation to experimental data, we believe that the study of the $\Delta I=1$ HPV on lattice should receive the same amount, if not more, of attention as the $\Delta I=2$ one. In this paper we investigate in some detail how a continuous EFT may help in the future lattice calculation of $h_\pi^1$.

In Ref.~\cite{Feng:2017iqb} we show that $h_\pi^1$ can be recast as a neutron-proton mass splitting induced by a set of $\Delta I=1$ P-even four-quark operators. Improving from the limitations of PCAC, {\color{black}we show by considering the long- and short-range higher-order corrections that such a relation holds with a precision better than $10\%$ even with a conservative estimation.}  This observation turns the lattice study of $h_\pi^1$ into computations of P-even three-point correlation functions involving only five sets (three independent combinations) of contractions. Two combinations among them are quark loop contractions which are in principle noisy, but in this work we show that one can obtain partial information of the chiral logarithms in the quark loop diagrams by studying the much easier connected diagrams. We further demonstrate that one only needs to perform a global fit with four  independent LECs, two of which can be obtained easily from connected diagrams, in order to completely determine the LO and chiral logarithmic terms in $h_\pi^1$ induced by non-strange operators. Approximate relations among LECs based on the spin-flavor symmetry are also derived to facilitate the global fit. For operators with strange quark fields, fitting of one single LEC from the computation of one quark loop contraction is needed to determine the LO and chiral logarithmic terms. We hope that the analysis above will provide extra motivations for the lattice community to perform an up-to-date first-principle computation of $h_\pi^1$ which will constitute a new breakthrough in our understanding of HWI.

\section*{Acknowledgements} 
The authors thank Jordy de Vries, Xu Feng and Liuming Liu for many inspiring discussions. We thank Ulf-G.~Mei{\ss}ner for a careful reading of this manuscript and for his useful comments. This work is supported in part by the National Natural Science Foundation of China (NSFC) under Grant Nos.~11575110, 11655002, 11735010 and 11747601, by NSFC and Deutsche Forschungsgemeinschaft (DFG) through funds provided to the Sino--German Collaborative Research Center ``Symmetries and the
Emergence of Structure in QCD'' (NSFC Grant No.~11621131001), by the Natural Science Foundation of Shanghai under Grant Nos.~15DZ2272100 and 15ZR1423100, by Shanghai Key Laboratory for Particle Physics and Cosmology, by the Key Laboratory for Particle Physics, Astrophysics and Cosmology, Ministry of Education,  by the CAS Key Research Program of Frontier Sciences (Grant No.~QYZDB-SSW-SYS013), by the CAS Key Research Program (Grant No.~XDPB09), and by the CAS Center for Excellence in Particle Physics (CCEPP). We also appreciate the supports through the Recruitment Program of Foreign Young Talents from the State Administration of Foreign Expert Affairs, China, and the Thousand Talents Plan for Young
Professionals.

\begin{appendix}
	
\section{\label{sec:Otherint}Other choices of the baryon interpolating operator}

Here we shall comment on our choice of baryon interpolator in Eq.~\eqref{eq:interpolator}. We choose this form because it satisfies both the exchange symmetries in Eq.~\eqref{eq:flavorsymmetry} required to single out the spin-1/2 baryons with the least number of terms. At the same time, we realize that in actual lattice calculations interpolators with simpler Dirac structures such as
\begin{eqnarray}
\chi_1(q_1,q_2,q_3)&=&\varepsilon^{abc}(q_1^{aT}C\gamma_5q_2^b)q_3^c , \nonumber\\
\chi_2(q_1,q_2,q_3)&=&\varepsilon^{abc}(q_1^{aT}Cq_2^b)\gamma_5q_3^c
\end{eqnarray} 
are more commonly used. The problem is that they do not satisfy the exchange symmetry relations in Eq.~\eqref{eq:flavorsymmetry} and are thus not sufficient to specify a spin-1/2 baryon. One may construct the linear combinations
\begin{equation}
\chi_i'(q_1,q_2,q_3)=\chi_i(q_1,q_2,q_3)+\chi_i(q_1,q_3,q_2)
\end{equation}
($i=1,2$) that do satisfy Eq.~\eqref{eq:flavorsymmetry}, but the price is that now each interpolator contains two terms instead of one. In fact, using Fierz identity one is able to show that
\begin{equation}
\chi(q_1,q_2,q_3)=\chi_1'(q_1,q_2,q_3)-\chi_2'(q_1,q_2,q_3) ,
\end{equation} 
so these choices of interpolators are not all independent. Furthermore, it is demonstrated that the interpolator $\chi_2'$ has negligible overlap with the ground-state baryon~\cite{Leinweber:1994nm}. The explanation is that $\chi_2'$ scales as $\mathcal{O}(p^2/E^2)$ in the non-relativistic expansion, which implies that it overlaps more with excited states than with the ground state~\cite{Fodor:2012gf}. Therefore, as far as this work is concerned, one could legitimately replace $\chi(q_1,q_2,q_3)\rightarrow \chi_1'(q_1,q_2,q_3)$ without affecting any of the discussions above. 

\section{\label{sec:MESU42}Contraction diagrams as SU(4$|$2) matrix elements}

In this appendix we demonstrate how each quantity $M_X$ can be expressed as a linear combination of four-quark matrix elements in SU(4$|$2). Such expressions are of course not unique. As a simple illustration, we fix the external state to be proton and choose different four-quark operators of which matrix elements are taken. One can then easily verify that
\begin{eqnarray}
M_{V_1}&=&\left\langle p\right|\bar{j}\Gamma^\mu j \bar{k}\Gamma_\mu k\left|p\right\rangle, \nonumber\\
M_{V_2}&=&\left\langle p\right|\bar{j}\Gamma^\mu j \bar{j}\Gamma_\mu j\left|p\right\rangle-M_{V_1}, \nonumber\\
M_{D_{a1}}&=&\left\langle p\right|\bar{u}\Gamma^\mu j \bar{j}\Gamma_\mu u\left|p\right\rangle-M_{V_2}, \nonumber\\
M_{D_{a2}}&=&2\left\langle p\right|\bar{d}\Gamma^\mu j \bar{j}\Gamma_\mu d\left|p\right\rangle-2M_{V_2}, \nonumber\\
M_{D_{b1}}&=&\left\langle p\right|\bar{u}\Gamma^\mu u \bar{j}\Gamma_\mu j\left|p\right\rangle-M_{V_1}, \nonumber\\
M_{D_{b2}}&=&2\left\langle p\right|\bar{d}\Gamma^\mu d \bar{j}\Gamma_\mu j\left|p\right\rangle-2M_{V_1}, \nonumber\\
M_{C_1}&=&\left\langle p\right|\bar{u}\Gamma^\mu u \bar{u}\Gamma_\mu u\left|p\right\rangle-2M_{D_{a1}}-2M_{D_{b1}}-M_{V_1}-M_{V_2}, \nonumber\\
M_{C_2}&=&\left\langle p\right|\bar{u}\Gamma^\mu u \bar{d}\Gamma_\mu d\left|p\right\rangle-M_{D_{b1}}-\frac{1}{2}M_{D_{b2}}-M_{V_1}.\label{eq:MXME}
\end{eqnarray} 
This also implies that each contraction function $X(\tau,\tau_0,\tau')$ is related to a linear combination of the ``physical'' matrix elements in SU(4$|2$) and thus possesses the correct asymptotic exponential behavior $\exp\{-m_N(\tau-\tau')\}$. Therefore, each $M_X$ can be obtained on lattice in the same way as how the full matrix element $\left\langle p\right|\hat{O}\left|p\right\rangle$ is obtained. 

An immediate application of Eq.~\eqref{eq:MXME} is the calculation of all $\{M_X\}$ at tree level using the LO Lagrangian in Eq.~\eqref{eq:L4qLO}. The results read
\begin{eqnarray}
M_{V_1}&=&8\alpha_{V_1} m_N , \nonumber\\
M_{V_2}&=&8\alpha_{V_2}m_N , \nonumber\\
M_{D_{a1}}&=&\frac{2}{3}\left(5\alpha_{D_{a1}}+2\alpha_{D_{a2}}\right)m_N , \nonumber\\
M_{D_{a2}}&=&\frac{4}{3}\left(\alpha_{D_{a1}}+4\alpha_{D_{a2}}\right)m_N , \nonumber\\
M_{D_{b1}}&=&\frac{2}{3}\left(5\alpha_{D_{b1}}+2\alpha_{D_{b2}}\right)m_N , \nonumber\\
M_{D_{b2}}&=&\frac{4}{3}\left(\alpha_{D_{b1}}+4\alpha_{D_{b2}}\right)m_N , \nonumber\\
M_{C_1}&=&\frac{4}{3}\left(4\alpha_{C_1}+\alpha_{C_2}\right)m_N , \nonumber\\
M_{C_2}&=&\frac{2}{3}\left(2\alpha_{C_1}+5\alpha_{C_2}\right)m_N,
\end{eqnarray}
which give clear meaning of each LEC in terms of contraction diagrams.

\section{Essentials of HB PQChPT\label{sec:HBPQxPT}}

In this appendix we summarize the basic results of HB PQChPT that are used in this paper.

\subsection{Grading factor}

In PQQCD there are both fermionic and bosonic quarks. To determine whether a pair of quark fields commute or anti-commute, it is convenient to define a quantity $\eta_i$ such that $\eta_i=1(0)$ when $i$ is a fermionic (bosonic) index. Then, any two quantities $A$ and $B$ are said to have grading factors
$\eta_{A}$ and $\eta_{B}$, respectively, if
\begin{equation}
AB=(-1)^{\eta_{A}\eta_{B}}BA.
\end{equation}
For instance, a quark field $q_{i}'$ obviously has a grading factor $\eta_i$ so a pair of quark fields anti-commute only when both of them are fermionic. Meanwhile, for a matrix $\Gamma$ in the flavor space, its matrix element $\Gamma_{ij}$ in general has a grading factor of $\eta_i+\eta_j$. 

\subsection{The pNG particles}

Similar to the ordinary ChPT, the pNG particles (we refrain from using the word ``boson" because they can be either bosonic or fermionic though the spin is always 0) are contained in the matrix $U$. However, here it is more preferable to parameterize $U$ as
\begin{equation}
U=\exp\left\{\frac{i\sqrt{2}\Phi}{F_0}\right\}
\end{equation}
and to study the propagator of $\Phi_{ij}$. By doing so we avoid the need to define explicitly all the $9N^2-1$ generators in a $\mathrm{SU}(2N|N)$ PQChPT. With all quarks degenerate, the pNG propagator takes the following compact form~\cite{Sharpe:2001fh},
\begin{equation}
\langle T\{ \Phi_{ij}\Phi_{j'i'}\}\rangle=\frac{i}{k^2-m_\pi^2+i\epsilon}\left[\delta_{ij}\delta_{i'j'}\left(\delta_{ii'}\varepsilon_i-\frac{1}{N}\right)+\left(1-\delta_{ij}\right)\delta_{ii'}\delta_{jj'}T_{ij}\right], \label{eq:Phiprop}
\end{equation}
where $\varepsilon_i=(-1)^{\eta_i+1}$, and $T_{ij}$ equals $-1$ when both ${i,j}$ are ghost indices and 1  otherwise. The first term at the RHS of Eq.~\eqref{eq:Phiprop} is contributed by the neutral particles while the second term is by the charged particles. 

\subsection{Three-index representation of spin-1/2 baryons}

The spin-half baryons in PQChPT are usually represented by a three-index
form $B_{ijk}^{\gamma}$, where $\{i,j,k\}$ are flavor indices and
$\gamma$ is the Dirac index~\cite{Labrenz:1996jy}. It can be most easily understood by comparing with a three-quark representation,
\begin{equation}
B_{ijk}^{\gamma}\sim\varepsilon^{abc}\left(C\gamma_{5}\right)_{\alpha\beta}\left[q_{i}^{\prime\alpha a}q_{j}^{\prime\beta b}q_{k}^{\prime\gamma c}-q_{i}^{\prime\alpha a}q_{j}^{\prime\gamma c}q_{k}^{\prime\beta b}\right].\label{eq:3quark}
\end{equation}
Of course in ChPT one does not deal explicitly with quark fields, but Eq.~\eqref{eq:3quark}
is still useful in determining the symmetries and transformation rules
of $B_{ijk}$ as follows.
\begin{itemize}
	\item Symmetries under the exchange of two flavor indices:
	\begin{eqnarray}
	B_{ijk} & = & (-1)^{\eta_{j}\eta_{k}+1}B_{ikj},\nonumber \\
	B_{ijk} & = & (-1)^{\eta_{i}\eta_{j}}B_{jik}+(-1)^{\eta_{i}\eta_{j}+\eta_{j}\eta_{k}+\eta_{k}\eta_{i}}B_{kji}, \label{eq:exchange}
	\end{eqnarray}
	which tell us: (1) $B_{iii}=0$; (2) if two out of three indices are the same,
	then there is only one independent field; (3) if all three indices
	are different, then there are only two independent fields. 
	\item Transformation of $B_{ijk}$ under the chiral rotation~\cite{Labrenz:1996jy}:
	\begin{equation}
	B_{ijk}\rightarrow(-1)^{\eta_{i'}(\eta_{j}+\eta_{j'})+(\eta_{k}+\eta_{k'})(\eta_{i'}+\eta_{j'})}K_{ii'}K_{jj'}K_{kk'}B_{i'j'k'},
    \label{eq:Bijkchiral}
	\end{equation}
	which can be understood by first going back to Eq.~\eqref{eq:3quark},
	making the transformation $q'\rightarrow Kq'$, and moving $K_{jj'},K_{kk'}$
	through the quark fields to the left. 
\end{itemize}
The barred quantity of the baryon field is defined as $\bar{B}_{kji}\equiv\overline{(B_{ijk})}$.
It obviously satisfies the exchange symmetry relations,
\begin{eqnarray}
\bar{B}_{kji} & = & (-1)^{\eta_{j}\eta_{k}+1}\bar{B}_{jki}, \nonumber \\
\bar{B}_{kji} & = & (-1)^{\eta_{i}\eta_{j}}\bar{B}_{kij}+(-1)^{\eta_{i}\eta_{j}+\eta_{j}\eta_{k}+\eta_{k}\eta_{i}}\bar{B}_{ijk} ,
\end{eqnarray}
and the transformation rule
\begin{equation}
\bar{B}_{kji}\rightarrow(-1)^{\eta_{i'}(\eta_{j}+\eta_{j'})+(\eta_{k}+\eta_{k'})(\eta_{i'}+\eta_{j'})}\bar{B}_{k'j'i'}K_{k'k}^{\dagger}K_{j'j}^{\dagger}K_{i'i}^{\dagger}.
\end{equation}

\subsection{Constructing chiral invariants}

Here we introduce all the remaining ingredients needed to construct the chirally-invariant Lagrangian in the baryon sector. The definitions of the vector connection $\Gamma_\mu$ and the
axial vector $u_\mu$ (in the absence of external sources)
\begin{eqnarray}
\Gamma_{\mu} & = & \frac{1}{2}\left(u^{\dagger}\partial_{\mu}u+u\partial_{\mu}u^{\dagger}\right)\nonumber \\
u_{\mu} & = & i\left(u^{\dagger}\partial_{\mu}u-u\partial_{\mu}u^{\dagger}\right)
\end{eqnarray}
where $u=\sqrt{U}$, are formally identical to those in the ordinary HBChPT. The chiral
covariant derivative on $B_{ijk}$ can thus be defined as~\cite{Labrenz:1996jy}
\begin{equation}
\mathcal{D}_{\mu}B_{ijk}=\partial_{\mu}B_{ijk}+\left(\Gamma_{\mu}\right)_{ii'}B_{i'jk}+(-1)^{(\eta_{j}+\eta_{j'})\eta_{i}}\left(\Gamma_{\mu}\right)_{jj'}B_{ij'k}+(-1)^{(\eta_{i}+\eta_{j})(\eta_{k}+\eta_{k'})}\left(\Gamma_{\mu}\right)_{kk'}B_{ijk'}.
\end{equation}
The appearance of the grading factors can be easily understood as dictated by Eq.~\eqref{eq:Bijkchiral}. For instance, $(-1)^{(\eta_{j}+\eta_{j'})\eta_{i}}$
is required for $\left(\Gamma_{\mu}\right)_{jj'}$ to pass through
the index $i$ in $B_{ij'k}$ in order to act on the index $j'$.

As far as the one-loop analysis of contraction diagrams in this work is concerned, the only strong interaction Lagrangian we need is the $\mathrm{SU}(4|2)$ HB PQChPT Lagrangian at LO~\cite{Labrenz:1996jy},
\begin{equation}
\mathcal{L}_{B}=\bar{B}_{kji}iv\cdot \mathcal{D}B_{ijk}+\rho\bar{B}_{kji}S^{\mu}\left(u_{\mu}\right)_{kk'}B_{ijk'}(-1)^{(\eta_i+\eta_j)(\eta_k+\eta_{k'})}+\beta\bar{B}_{kji}S^{\mu}\left(u_{\mu}\right)_{ii'}B_{i'jk}.\label{eq:LBLO}
\end{equation}
One notices that there are two independent axial couplings $\rho$ and $\beta$, while there is only one $g_0$ in the LO SU(2) HBChPT Lagrangian. However, there are two axial couplings $D$ and $F$ in the SU(3) version (see, e.g., Ref.~\cite{Bernard:2007zu}), 
\begin{equation}
\mathcal{L}_{\rm SU(3)}= {\rm Tr}\left[\bar{B}_viv\cdot\mathcal{D}B_v\right]+ D\,{\rm Tr}\left[\bar{B}_v S^\mu \{u_\mu,B_v\}\right]+ F\,{\rm Tr}\left[\bar{B}_vS^\mu [u_\mu,B_v] \right] ,
\end{equation}
where $B_v$ denotes the baryon octet in Eq.~\eqref{eq:Boctet} and $u_\mu$ takes the SU(3) form, with $D\approx0.81$ and $F\approx0.46$ and $g_0=D+F$.
By matching the $\pi^0$ and $\eta$ coupling terms for the nucleons, we find the following relations between the PQChPT couplings $\{\rho,\beta\}$ and the SU(3) axial couplings $\{D,F\}$~\cite{Labrenz:1996jy,Chen:2001yi} 
\begin{equation}
\rho= 2F+\frac23 D,\qquad 
\beta=F-\frac{5}{3}D,
\end{equation}
which reproduce the relations in Ref.~\cite{Beane:2002vq} ($g_0$ is written as $g_A$ therein)
\begin{equation}
\rho=\frac{4}{3}g_0+\frac{1}{3}g_1,\:\:\:\beta=\frac{2}{3}g_1-\frac{1}{3}g_0.
\end{equation}
once we identify $g_0=D+F$ and $g_1=2(F-D)$.


\subsection{Independent baryon fields and the baryon propagator}

The three-index baryon fields $\{B_{ijk}\}$ are not all independent due to the exchange symmetry relations in Eq.~\eqref{eq:exchange}. Thus, let us
denote the independent baryon fields as $\{B_{a}\}$, and $\{B_{ijk}\}$
can be expressed in terms of the independent fields as
\begin{equation}
B_{ijk}=\psi_{ijk}^{a*}B_{a} , \label{eq:BijkBa}
\end{equation}
where $\{\psi_{ijk}^{a}\}$ are c-numbers, and the complex conjugate is
just a convention. In fact, the equation above defines the coefficients
$\{\psi_{ijk}^{a}\}$, and some useful examples of these coefficients are summarized in Appendix~\ref{sec:psiijk}. As shown in Eq.~\eqref{eq:LBLO}, $B_{ijk}$ is normalized such that $\bar{B}_{kji}iv\cdot \mathcal{D}B_{ijk}$
reproduces the properly-normalized kinetic terms of each independent
baryon. This imposes the following orthonormal condition to the coefficients,
\begin{equation}
\psi_{ijk}^{a}\psi_{ijk}^{b*}=\delta_{ab}.
\end{equation}
The inversion of Eq. (\ref{eq:BijkBa}) is not unique; however, the
most convenient form of inversion is simply
\begin{equation}
B_{a}=\psi_{ijk}^{a}B_{ijk} , \label{eq:BaBijk}
\end{equation}
which is a direct consequence of the orthonormal condition. 

The baryon propagator is most conveniently expressed in terms of the
non-independent fields $\{B_{ijk}\}$,
\begin{equation}
\left\langle T\left\{B_{lmn}\bar{B}_{kji}\right\}\right\rangle =\frac{i}{v\cdot k+i\epsilon}F_{lmn,ijk}^{1/2} , \label{eq:prop}
\end{equation}
where\footnote{The expression of $F_{lmn,ijk}^{1/2}$ in Ref. \cite{Labrenz:1996jy} contains a couple of typos which are corrected here.}
\begin{eqnarray}
F_{lmn,ijk}^{1/2} & = & \frac{1}{6}\left(2\delta_{ijk}^{lmn}-2(-1)^{\eta_{j}\eta_{k}}\delta_{ikj}^{lmn}+(-1)^{\eta_{i}\eta_{j}}\delta_{jik}^{lmn}-(-1)^{\eta_{i}(\eta_{j}+\eta_{k})}\delta_{jki}^{lmn}\right.\nonumber \\
&  & \left.-(-1)^{\eta_{k}(\eta_{i}+\eta_{j})}\delta_{kij}^{lmn}+(-1)^{\eta_{i}\eta_{j}+\eta_{i}\eta_{k}+\eta_{j}\eta_{k}}\delta_{kji}^{lmn}\right)
\end{eqnarray}
with the shorthand $\delta_{ijk}^{lmn}\equiv\delta_{il}\delta_{jm}\delta_{kn}$.
Notice that the three flavor indices of the initial and final baryons could
be different as they may still represent the same independent baryon
field. In fact, since 
\begin{equation}
\left\langle B_{lmn}\bar{B}_{kji}\right\rangle =\psi_{lmn}^{a*}\psi_{ijk}^{b}\left\langle B_{a}\bar{B}_{b}\right\rangle =\frac{i}{v\cdot k+i\epsilon}\psi_{lmn}^{a*}\psi_{ijk}^{a}=\frac{i}{v\cdot k+i\epsilon}F_{lmn,ijk}^{1/2},
\end{equation}
we see that the factor $F_{lmn,ijk}^{1/2}=\psi_{lmn}^{a*}\psi_{ijk}^{a}$ plays the role
of projecting out the independent baryon fields. 

\subsection{Prescription for the correct usage of Feynman rules}

We shall name the Feynman vertices directly extracted from
the HB PQChPT Lagrangian as the ``na\"{\i}ve vertices'', which are in terms
of the non-independent baryon fields $\{B_{ijk}\}$. For instance,
the Lagrangian $\mathcal{L}=\lambda\bar{B}_{kji}\Phi_{ii'}B_{i'jk}$
gives the following na\"{\i}ve vertex: $i\mathcal{M}(B_{ijk}\Phi_{pq}\rightarrow B_{lmn})=i\lambda\delta_{iq}\delta_{lp}\delta_{jm}\delta_{kn}$.
Similarly, the propagator in Eq.~\eqref{eq:prop} should be known
as the ``na\"{\i}ve propagator''. 

A direct application of na\"{\i}ve Feynman propagators and vertices in the computation of physical amplitudes is obviously inappropriate. However, there is a simple prescription that ensures
the correctness of the final amplitude. For the simplicity of notation
we shall introduce another shorthand: We use Greek letters $\{\alpha\}$
to collectively represent the three indices $\{i,j,k\}$ in $B_{ijk}$;
meanwhile, Roman letters $\{a\}$ denote the independent baryon fields.
For instance, Eqs.~\eqref{eq:BijkBa} and \eqref{eq:BaBijk} now read
$B_{\alpha}^{}=\psi_{\alpha}^{a*}B_{a}^{}$ and $B_{a}^{}=\psi_{\alpha}^{a}B_{\alpha}^{}$,
respectively. The prescription is as follows:\\
To calculate an amplitude $i\mathcal{M}(B_{a_{1}}...B_{a_{n}}X\rightarrow B_{b_{1}}...B_{b_{n}}X')$
where $\{B_{a_{i}},B_{b_{j}}\}$ are physical baryon fields and $\{X,X'\}$
collectively represent all non-baryon fields, we shall first calculate
the more general amplitude
\begin{equation}
i\tilde{\mathcal{M}}_{\beta_{1}...\beta_{n},\alpha_{1}...\alpha_{n}}\equiv i\tilde{\mathcal{M}}(B_{\alpha_{1}}...B_{\alpha_{n}}X\rightarrow B_{\beta_{1}}...B_{\beta_{n}}X')
\end{equation}
without the external spinors using the na\"{\i}ve propagators and vertices. The actual amplitude $i\mathcal{M}$ is then obtained as
\begin{equation}
i\mathcal{M}=\psi_{\beta_{1}}^{b_{1}}...\psi_{\beta_{n}}^{b_{n}}\bar{u}_{b_{1}}...\bar{u}_{b_{n}}i\tilde{\mathcal{M}}_{\beta_{1}...\beta_{n},\alpha_{1}...\alpha_{n}}u_{a_{1}}...u_{a_{n}}\psi_{\alpha_{1}}^{a_{1}*}...\psi_{\alpha_{n}}^{a_{n}*} ,
\end{equation}
where $\{\alpha_{i},\beta_{j}\}$ are summed over but $\{a_{i},b_{j}\}$
are not. One can show that the contraction of $\psi,\psi^{*}$ to
the external states is equivalent to taking only the independent
baryonic DOFs in the loop calculation. A good thing
about this relation is that we now need only to know the coefficients
$\{\psi_{ijk}^{a}\}$ for ordinary baryons that exist as external
states, i.e., baryons consist of three fermionic (dynamical or valence) quarks.

\subsection{$\{\psi_{ijk}^{a}\}$ in SU(3)\label{sec:psiijk}}

In a theory with only three dynamical quarks $(q=u,d,s)$, the three-index
representation $B_{ijk}$ can be directly mapped to the baryon octet
as~\cite{Labrenz:1996jy,Chen:2001yi}
\begin{equation}
B_{ijk}=\frac{1}{\sqrt{6}}\left(\varepsilon_{ijm}B_{km}+\varepsilon_{ikm}B_{jm}\right) ,
\end{equation}
where the baryon octet matrix $B$ is defined as
\begin{equation}
B=\left(\begin{array}{ccc}
\frac{1}{\sqrt{6}}\Lambda+\frac{1}{\sqrt{2}}\Sigma^{0} & \Sigma^{+} & p\\
\Sigma^{-} & \frac{1}{\sqrt{6}}\Lambda-\frac{1}{\sqrt{2}}\Sigma^{0} & n\\
\Xi^{-} & \Xi^{0} & -\frac{2}{\sqrt{6}}\Lambda
\end{array}\right).
\label{eq:Boctet}
\end{equation}
With this we can easily obtain the independent non-vanishing coefficients
$\psi_{ijk}^{a}$ in SU(3):
\begin{equation}
\psi_{112}^{p}=-\psi_{221}^{n}=-\psi_{113}^{\Sigma^{+}}=\psi_{331}^{\Xi^{0}}=\psi_{223}^{\Sigma^{-}}=-\psi_{332}^{\Xi^{-}}=\frac{1}{\sqrt{6}},\psi_{123}^{\Lambda}=-\psi_{213}^{\Lambda}=-\frac{1}{2},\psi_{123}^{\Sigma^{0}}=\psi_{213}^{\Sigma^{0}}=\frac{1}{2\sqrt{3}}.
\end{equation}
All other non-vanishing coefficients can be obtained by symmetry:
$\psi_{ijk}^{a}=\psi_{ikj}^{a}$ and $\psi_{ijk}^{a}=-\psi_{jik}^{a}-\psi_{kji}^{a}$
(no grading factors because all quarks are fermionic). 

The results above can also be used to determine the coefficients $\psi$ for the other baryons in PQChPT. For example, the SU$(4|2)$ baryon states $\tilde{\Sigma}^0$ and $\tilde{\Lambda}$ we introduced in Section~\ref{sec:PQChPTanalysis} have the following independent non-vanishing coefficients:
\begin{equation}
\psi^{\tilde{\Sigma}^0}_{134}=\psi^{\tilde{\Sigma}^0}_{314}=\frac{1}{2\sqrt{3}},\:\:\psi^{\tilde{\Lambda}}_{134}=-\psi^{\tilde{\Lambda}}_{314}=-\frac{1}{2}.
\end{equation}

\end{appendix}



\bibliographystyle{JHEP-2}
\bibliography{hpipqxpt_ref}

\end{document}